\documentclass[aps,twocolumn,showlabels,showrefs,amsmath,amssymb,pre,superscriptaddress,floatfix,colors]{revtex4-1}

\pdfoutput=1

\usepackage{graphicx}
\usepackage{dcolumn}
\usepackage{bm}
\usepackage{amssymb}
\usepackage{multirow}
\usepackage{color}
\usepackage{hyperref}
\usepackage{ulem}
\usepackage{float}

\begin{document}

\title{Full phase diagram of active Brownian disks: \\
from melting to motility-induced phase separation}

\date{\today}


\author{Pasquale Digregorio}
\affiliation{Dipartimento  di  Fisica,  Universit\`a  degli  Studi  di  Bari  and  INFN,
Sezione  di  Bari,  via  Amendola  173,  Bari,  I-70126,  Italy}

\author{Demian Levis}
\affiliation{CECAM  Centre  Europ\'eeen  de  Calcul  Atomique  et  Mol\'eculaire,
Ecole  Polytechnique  F\'ed\'erale  de  Lausanne,  Batochimie,  Avenue  Forel  2,  1015  Lausanne,  Switzerland}
\affiliation{UBICS  University  of  Barcelona  Institute  of  Complex  Systems,  Mart\'{\i}  i  Franqu\`es  1,  E08028  Barcelona,  Spain}

\author{Antonio Suma}
\affiliation{SISSA - Scuola Internazionale Superiore di Studi Avanzati, Via Bonomea 265, 34136 Trieste, Italy}
\affiliation{Institute for Computational Molecular Science, Temple University, Philadelphia, PA 19122, USA }

\author{Leticia F. Cugliandolo}
\affiliation{Sorbonne Universit\'e, Laboratoire de Physique Th\'eorique et Hautes Energies, CNRS UMR 7589, 
4 Place Jussieu, 75252 Paris Cedex 05, France}

\author{Giuseppe Gonnella}
\affiliation{Dipartimento di Fisica, Universit\`a degli Studi di Bari and INFN,
Sezione di Bari, via Amendola 173, Bari, I-70126, Italy}

\author{Ignacio Pagonabarraga} 
\affiliation{CECAM  Centre  Europ\'eeen  de  Calcul  Atomique  et  Mol\'eculaire,
Ecole  Polytechnique  F\'ed\'erale  de  Lausanne,  Batochimie,  Avenue  Forel  2,  1015  Lausanne,  Switzerland}
\affiliation{UBICS  University  of  Barcelona  Institute  of  Complex  Systems,  Mart\'{\i}  i  Franqu\`es  1,  E08028  Barcelona,  Spain}

\begin{abstract}
We establish the complete phase diagram of self-propelled hard disks in two spatial dimensions 
from the analysis of the equation of state and the statistics of local order parameters.
The equilibrium melting scenario is maintained at small activities, with coexistence between active liquid 
and hexatic order, followed by a proper hexatic phase and a further transition to an active solid.  As activity increases,  the emergence of 
hexatic and solid order is shifted towards higher densities. Above a critical activity  and for a certain range of packing fractions, the system undergoes MIPS and demixes into low and 
high density phases; the latter can be either disordered (liquid) or ordered (hexatic or solid) depending on   activity.
\end{abstract}

\maketitle

Active materials are out-of-equilibrium systems in which the dynamics of their elements
 break detailed balance \cite{MarchettiRev}.   
Examples can be found  
 in living systems, {\it e.g.} the collective motion of large animal groups~\cite{Couzin2003, Bialek2012}, bacteria swarming~\cite{Zhang2010}, 
 and the formation of traveling fronts of actin filaments~\cite{Schaller2010}, as well as in synthetic ones, like self-propelled grains~\cite{Deseigne2010} or 
 self-catalytic colloidal suspensions~\cite{BechingerRev}.
Despite such diversity, the emergence of activity-induced collective behavior is
captured by minimal models that yield accurate descriptions and shed light on their universal character. 
A key example is
the Active Brownian Particles (ABP) model   which  considers spherical self-propelled 
particles with only excluded volume interactions~\cite{Hagen2011,Romanczuk2012, Fily2012, Redner2013f, CatesTailleur2013, Bialke2013}. 
A hallmark of  active particle systems is that 
at high enough density and activity, self-propulsion triggers a motility-induced phase separation (MIPS) into a low-density gas  in coexistence with a 
high-density drop~\cite{TailleurCates2008, Fily2012, Redner2013f, CatesTailleur2013, Bialke2013, Stenhammar2014, Bialke2015, CatesRev}, 
resembling the equilibrium liquid-gas transition but in 
the absence of cohesive forces and  without a thermodynamic support~\cite{Levis2017a, Solon2018thermo}. 

Although active particles can in principle move in 3D, in most experimental set-ups they are confined to 2D.
Most studies of 2D ABP focused on MIPS, 
and have therefore  been largely restricted to  intermediate 
densities~\cite{TailleurCates2008, Fily2012, Redner2013f, CatesTailleur2013, Bialke2013, Stenhammar2014, Bialke2015, CatesRev,Levis2017a, Solon2018thermo}.
In contrast, their solidification, or melting,  has received little attention {~\cite{Bialke2012cryst, Briand2016, KKK}}, and the 
connection between the high Pe behavior and the equilibrium physics 
as $\text{Pe}\to0$ has been, surprisingly, disregarded. 
In particular, the fate of 2D melting (with its intermediate hexatic phase) under active forces, has {been} investigated for 
dumbbell systems~\cite{Cugliandolo2017}, where MIPS is continuously connected to the passive liquid-hexatic coexistence. This result sheded new light on 
the very nature of  MIPS and showed the importance of exploring the full phase diagram at high densities. In this Letter, we address this issue in the paradigmatic
ABP model.  
 
Melting in 2D is a fundamental problem that has remained elusive despite decades of intensive research~\cite{Strandburg1988,Ryzhov2017}.
The transition was initially claimed to be first order~\cite{Alder1962} and later argued to follow 
a different scenario, with an intermediate hexatic phase, separated by continuous transitions mediated by the 
unbinding of defects~\cite{Kosterlitz1973, Halperin1978, Young1979}. More recently,  numerical simulations \cite{Bernard2009a,Bernard2011g,Kapfer2015b} followed by experiments on colloidal monolayers \cite{Dullens2017}, clarified the picture.  
They indicate that melting of passive hard-disks takes place in two steps: as the packing fraction is increased, a first-order transition between 
the liquid and  hexatic phases occurs, followed by a continuous Berezinskii-Kosterlitz-Thouless (BKT) transition between the hexatic and the solid. The hexatic 
phase exhibits quasi-long-range orientational order and short-range positional one, while the solid phase has quasi-long-range positional and long-range 
orientational order. Liquid and hexatic phases coexist close to the liquid phase, within a narrow interval of packing fractions.

\begin{figure}[h]
\centering 
\includegraphics[scale=0.12]{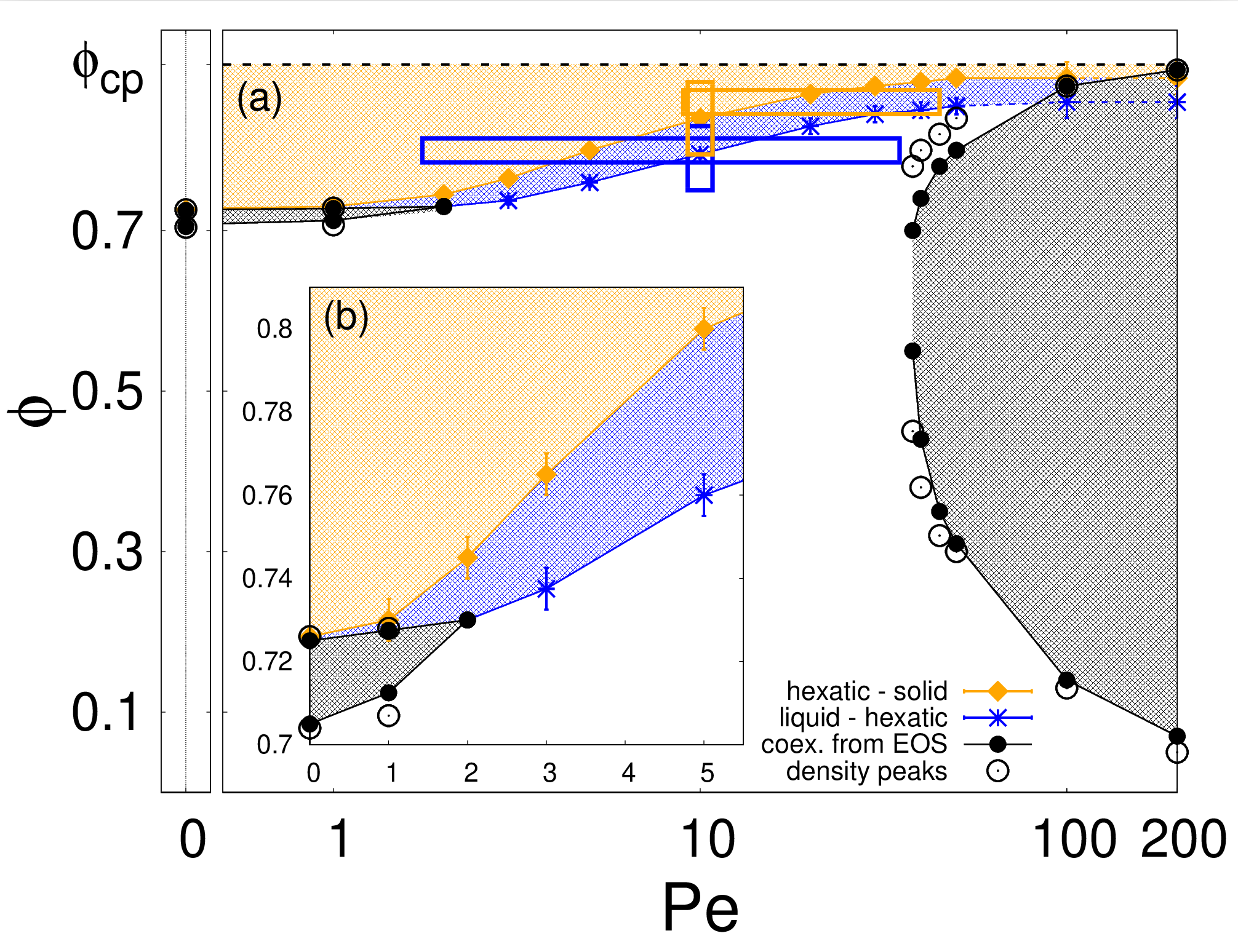}   
\caption{$\text{Pe}$-$\phi$ phase diagram of ABP. In the inset, a zoom over the low Pe - high $\phi$ regime highlights the connection with 2D melting. 
In the black area there is coexistence, in the blue hexatic order and the orange one is an active solid phase.
The black filled (open) points were obtained from pressure (density distribution) measurements; the
blue stars from the orientational correlations and the orange symbols from the spatial correlations decay.  
Although  narrow,  the blue area persists for $\text{Pe}\to 0$ and broadens as Pe increases. 
The {solid} boxes highlight  the parameters used to measure the correlations in Figs.~\ref{fig:g6} and \ref{fig:positional}.
}
\label{fig:phase-diagram}
\end{figure}

Here we examine how activity affects the phase behavior of  2D systems of isotropic particles {(i.e. with no alignment interactions)},  
from the dilute regime  to close packing ($\phi_{\rm cp} \approx 0.91$). We establish the complete phase 
diagram of 2D ABP  spanning a broad range of activities, see Fig.~\ref{fig:phase-diagram}.
We show that the two-step melting scenario at Pe = 0 is maintained  at finite but small activity,  
with a {\it coexistence region between active liquid and hexatic phases}  (black area). 
Above, an {\it active hexatic phase} exists for all the explored activities (blue sector). Strikingly, active disks arrange in a 
hexatic phase in a larger density range than passive ones. 
At higher densities, orientational long-range and positional quasi-long-range order emerge for any activity, signaling the presence of an {\it active solid phase}
(orange region). The  liquid-hexatic  and  hexatic-solid transitions shift towards higher densities with Pe, 
meaning that activity destabilizes the ordered phases. 
At high enough activity ($\text{Pe}\gtrsim35$), we identify the boundaries of MIPS  using both pressure measurements and density distributions 
(black and white symbols). 
The MIPS region broadens as activity increases and eventually crosses the hexatic and solid transition lines. Such results  
show that (i) MIPS prevails over the hexatic and solid phases and (ii) 
MIPS  generates a phase separation between a dilute  and a high-density phase, which can  either be liquid,  hexatic or solid, as activity is increased.

We consider $N$ overdamped ABP, in a square box with volume $V=L^2$ and periodic boundary conditions. They self-propel 
under a constant  modulus force $F_{\rm act}$ along $\bold{n}_i=( \cos{\theta_i(t)},\sin{\theta_i(t)})$  and obey
\begin{equation}
	\label{eq:langevin}
	\gamma\dot{\bold{r}}_i=F_{\rm act} \bold{n}_i- {\boldsymbol{\nabla}}_i\sum_{j(\neq i)}U(r_{ij}) + \bm{\xi}_i \; ,\quad \dot{\theta}_i=\eta_i \; , 
	\end{equation}
with ${\bold{r}}_i$ the position of the center of the $i$th particle, $r_{ij}=|{\bold{r}}_i-{\bold{r}}_j|$ the inter-particle distance, and a short-ranged repulsive potential, 
$U(r)=4\varepsilon [({\sigma}/{r})^{64}-({\sigma}/{r})^{32}]+\varepsilon$ if $r< \sigma_d=2^{1/32}\sigma$ and $0$ otherwise.
The terms $\bm{\xi}$ and $\eta$ are zero-mean  Gaussian noises that verify 
$\langle \bm{\xi}_{i}(t) \, \bm{\xi}_{j}(t') \rangle = 2 \gamma k_B T \delta_{ij} \delta(t-t') \bold{1}$ and $\langle \eta_{i}(t) \, \eta_{j}(t') \rangle = 2 D_{\theta} \delta_{ij} \delta(t-t')$. 
The units of length, time and energy are given by  $ \sigma_d$, $\tau= D^{-1}_\theta$ and $\varepsilon$, respectively. 
We fix $D_\theta = 3\gamma k_BT/{\sigma^2_d}$ and vary  the packing fraction 
$\phi =\pi{\sigma^2_d}N/(4V)$   and P\'eclet number Pe = $F_{\rm act} {\sigma_d}/(k_BT)$ by tuning $L$ and $F_{\rm act}$ at fixed $\gamma=10$ and $k_BT=0.05$.  
The integration of Eqs.~(\ref{eq:langevin}) used the velocity Verlet algorithm implemented in LAMMPS~\cite{PlimptonLAMMPS,myfoot}
Simulations ran with $N=256^2$ particles, scanning the parameter space  $\phi\in[0:0.9]$ and $\text{Pe}\in[0:200]$. With 
less ($N=128^2$) and more ($N=512^2$) particles we 
explored finite size effects.

{\it The equation of state.}
Our  first estimate of the phase boundaries is given by the $\phi$ dependence of the mechanical
pressure~\cite{Winkler2015, Levis2017a} 
\begin{equation}
	\Delta P 
	\! = \!
\frac{F_{\rm act}}{2V} \! \sum_i\langle \bold{n}_i \cdot \bold{r}_i \rangle 
	- \! \frac{1}{4V} \! \sum_{i,j} \langle  {\boldsymbol{\nabla}}_i U(r_{ij}) \! \cdot \! (\bold{r}_i-\bold{r}_j) \rangle
\label{eq:pressure}
\end{equation}
with $\Delta P = P-P_G$ and $P_G= N k_BT/V$ the ideal gas pressure. 
The first term,  $P_{\rm act}$, quantifies the effect of $F_{\rm act}$, the so-called 
active or swim pressure \cite{Brady2014, Solon2015a}. The second one, $P_{\rm int}$, is the standard virial 
term due to particle interactions. 
The definition in Eq.~(\ref{eq:pressure}) 
is a state function for isotropic ABP such that $P(\phi)$ defines an equation of state~\cite{Solon2015a}. 
(This does not hold generically in active systems for which the pressure can, for instance, depend on the details of the interaction  between 
the particles and the confining walls~\cite{Solon2015}.) In the dilute limit we recover the ideal gas law 
$
	PV=N k_BT_{\rm eff}= Nk_BT (1+\mbox{Pe}^2/6)
$,
at an effective temperature that is compatible with the one that stems from the fluctuation-dissipation relation in the late diffusive 
regime~\cite{Suma2014, Levis2015, Ginot2015}. 

\begin{figure}[h!]
\vspace{0.2cm}
\centering 
\includegraphics[scale=0.13]{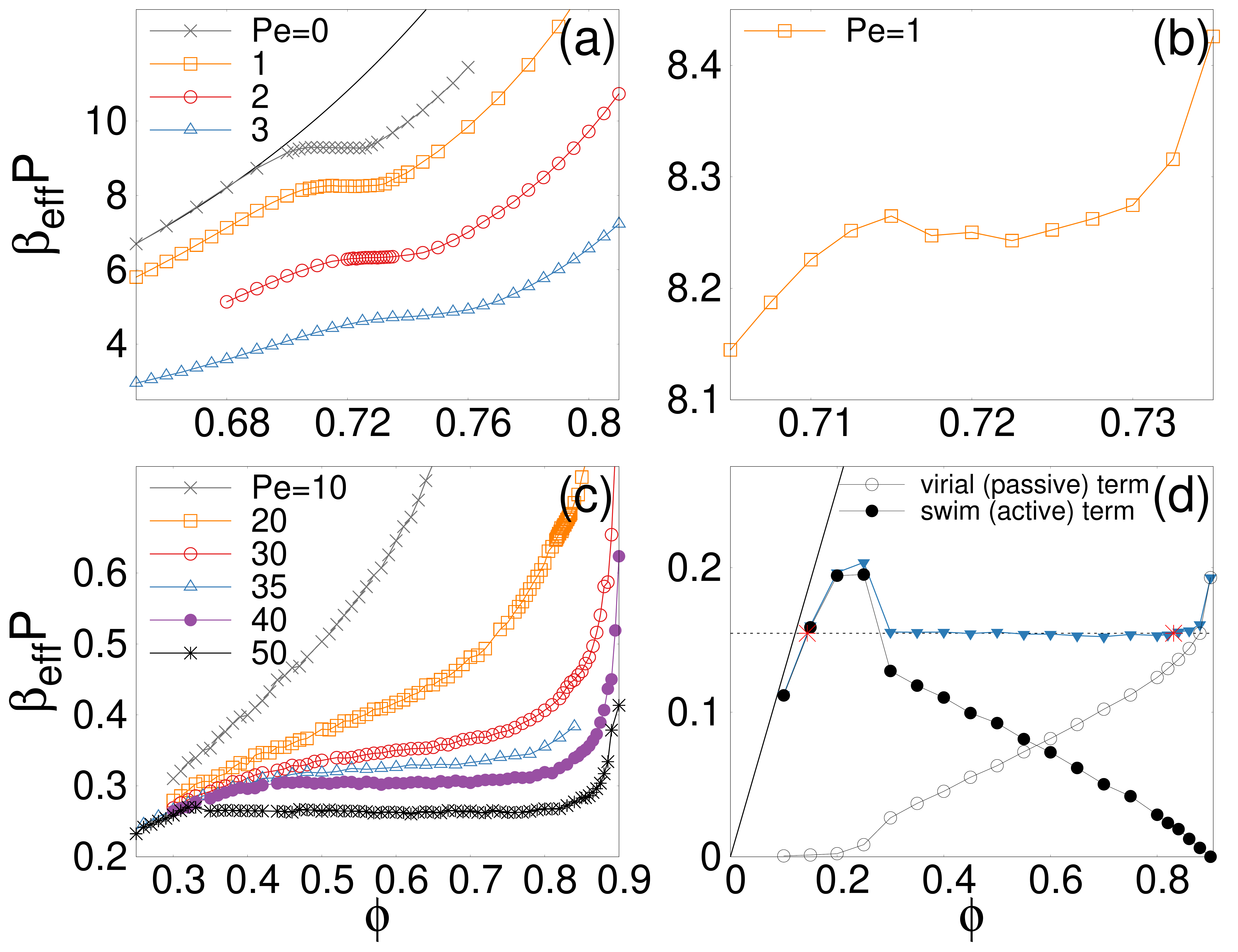}
\vspace{-0.45cm}
\caption{Equation of state.
(a) Numerical data for low Pe and analytical form for passive hard disks (continuous line)~\cite{kolafa2006}. 
(b) Details around the liquid-hexatic coexistence region  for $\text{Pe}=1$.
(c) Data for intermediate Pe.
(d) Swim $P_{\rm act}$ and interaction $P_{\rm int}$ contributions to $P$ in Eq.~(\ref{eq:pressure}) from the gas to the solid  at 
Pe = 100.  {Triangular symbols correspond to the total pressure.} The ideal gas law 
is shown with a continuous line. The red symbols indicate the coexistence densities reported in Fig. \ref{fig:phase-diagram}.
}
\label{fig:pressure}
\end{figure}

The equation of state  for zero and weak Pe is shown in Fig.~\ref{fig:pressure} (a).   
$P(\phi)$ is roughly flat in a narrow $\phi$ interval for $\text{Pe}\lesssim 3$. 
A zoom over  this area in the  $\text{Pe}=1$ case evidences a double loop 
structure characteristic of phase coexistence, 
see Fig.~\ref{fig:pressure} (b). 
Although the equal-area Maxwell construction, that allows to directly extract the binodals, 
cannot be readily applied  for $\text{Pe}>0$ \cite{Solon2015a, Solon2018thermo}, we  use
it by extension of the passive disks analysis~\cite{Bernard2011g}, as a first identification of the coexistence region  
(black dots in  Fig.~\ref{fig:phase-diagram}). 
Beyond $\text{Pe}=3$,
we do not find evidence for coexistence until the high-Pe regime where MIPS  is attained. For $\text{Pe}\gtrsim 35$ the  $P(\phi)$ curves 
become flat in between two densities.  
 Representative curves at $10 \leq$  Pe $\leq 50$ are displayed in Fig.~\ref{fig:pressure} (c). 
As it has been  recently reported~\cite{Winkler2015, Levis2017a}, at very high Pe, the pressure drops abruptly at the vicinity of MIPS (see Fig.~\ref{fig:pressure} (d)), 
as a consequence of the existence of a metastability region with a very large nucleation 
barrier~\cite{Levis2017a}. We obtain the limits of MIPS with an extrapolation of  the flat part of $P(\phi)$ across the pressure jump (or spinodal), as illustrated 
in Fig.~\ref{fig:pressure} (d) for  $\text{Pe}=100$. 
Previous numerical studies used the local density probability distribution functions (PDF) to locate 
the MIPS region, see {\it e.g.}~\cite{Redner2013f, Cugliandolo2017}. 
For the sake of completeness, we searched for the limits of a double peak structure of these PDFs, 
finding the open symbols in  Fig.~\ref{fig:phase-diagram}, 
in very good agreement with the pressure measurements (see Figs.~S1, S2, S3 in the Supplementary Material~\cite{SM} for further details).

\begin{figure}[h!]
\vspace{0.1cm}
\begin{center}
\includegraphics[scale=0.13,angle=0]{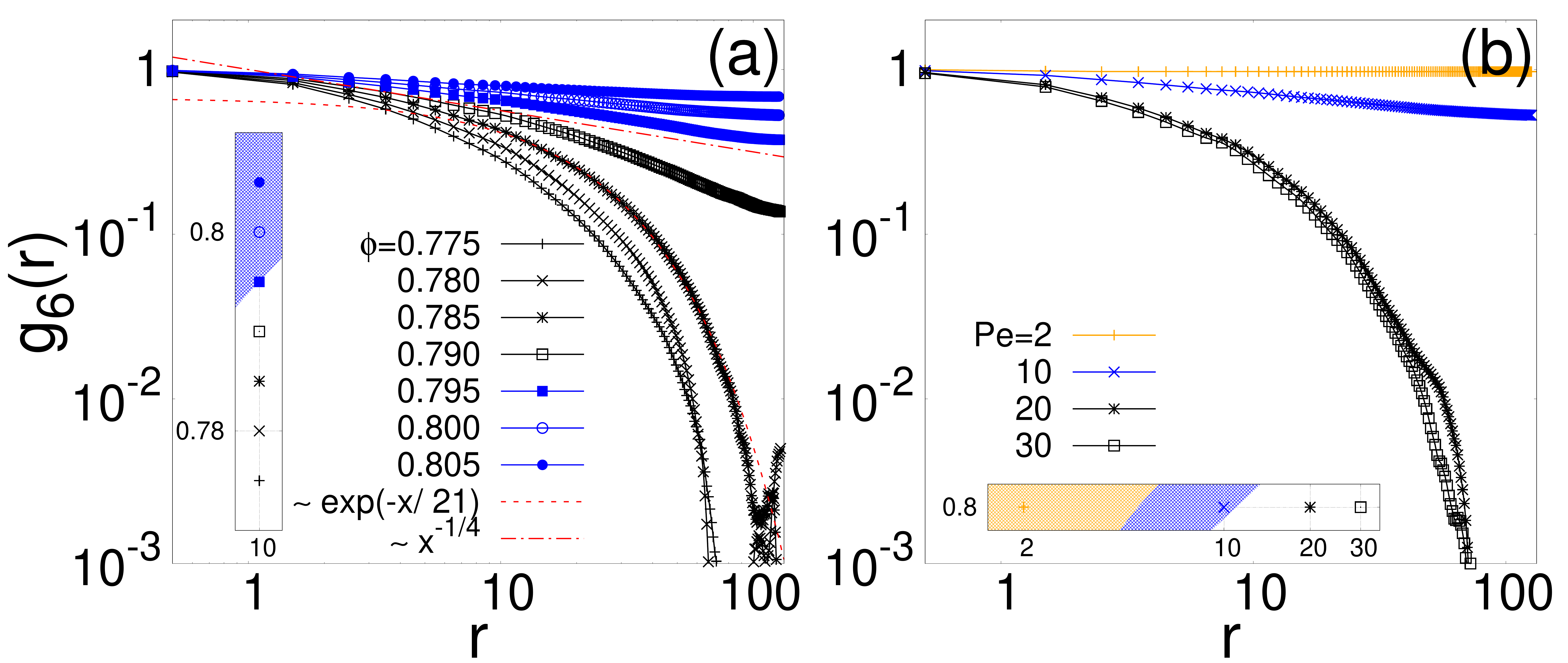}
\end{center}
\vspace{-0.45cm}
\caption{Orientational correlations  close to the active liquid-hexatic transition for 
varying densities at Pe = 10 (a) and different Pe  
at $\phi=0.80$ (b). The vertical and horizontal boxes represent the location of the parameters in  the phase diagram  
(see Fig.~\ref{fig:phase-diagram}).   
An exponential fit to the liquid data and an algebraic decay with power $\eta=1/4$ are shown in (a). 
}
\label{fig:g6}
\end{figure}

\begin{figure}[h!]
\vspace{0.3cm}
\begin{center} 
\includegraphics[scale=0.13]{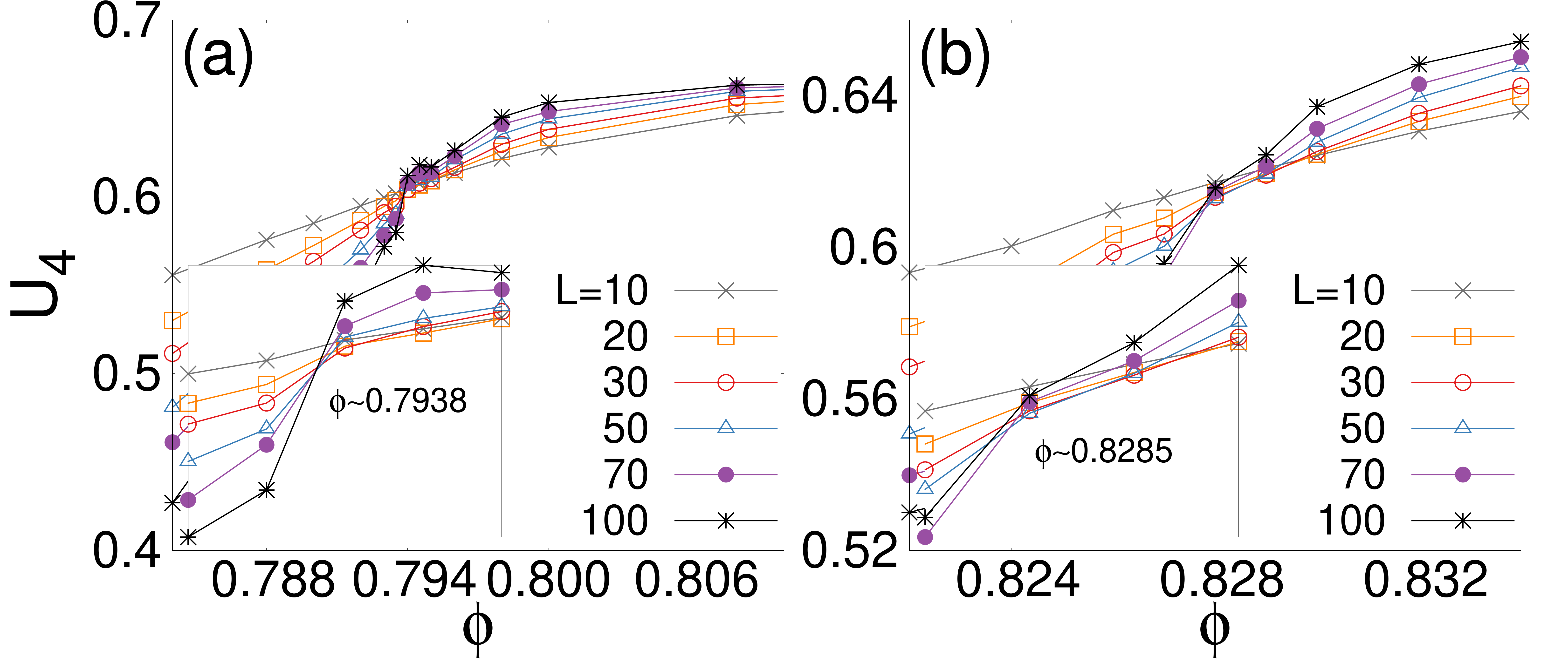}
\end{center}
\vspace{-0.45cm}
\caption{The system size dependence of the Binder cumulant  
for Pe = 10 (a) and Pe = 20 (b). 
}
\label{fig:Binder}
\end{figure}

{\it Orientational order and the hexatic phase.}
We put the orientational order to the test using the hexatic order parameter 
  {$\psi_6({\bold r}_j) = N_{ j}^{-1} \sum_{k=1}^{N_{ j}} e^{{\rm i} 6\theta_{jk}}$}, where $\theta_{jk}$ is the angle formed by
the segment that connects the center of the $j$th disk and the one of its $k$th (out of $N_j$) nearest neighbor 
(found with a Voronoi tessellation algorithm) and the $x$ axis. We studied its correlation function $g_6(r=|{\bold r}_j - {\bold r}_k|) 
= \langle \psi_6({\bold r}_j) \psi_6({\bold r}_k)\rangle/\langle \psi_6^2({\bold r}_j)\rangle$ 
and kurtosis or Binder parameter $U_4=1-\langle \psi^4_6({\bold r}_j) \rangle/(3\langle \psi^2_6({\bold r}_j) \rangle^2)$,  
see Figs.~\ref{fig:g6} and \ref{fig:Binder}, respectively. 
{ We use the change of behavior of $g_6(r)$, from exponential (active liquid, in black) to algebraic
$r^{-\eta}$ (active hexatic, in blue), as a criterion to locate the hexatic transition (blue symbols in Fig.~\ref{fig:phase-diagram})}.
{In the hexatic (blue) region the power law decay is maintained, with exponent $\eta$ taking a value close to the BKT  $\eta=1/4$ at the transition but varying with $\phi$ and Pe.} 
These data are compatible with the behavior of the Binder cumulant, $U_4$, that in
the scale of the main panel in Fig.~\ref{fig:Binder} has a common intersection point, proving the 
 transition. The zoom in the insert shows a weak remanent $N$-dependence that would be 
compatible with a first order phase transition~\cite{Vollmayr1993,Weber1995a}; however, the accuracy of our data is not enough to 
draw such a conclusion and, moreover, a second order transition is  consistent with the  absence of phase coexistence found above $\text{Pe}\approx 3$.
As illustrated in Fig.~\ref{fig:g6} (b), activity shifts the emergence of 
orientational quasi-long-range order to higher densities. 

\begin{figure}[h!]
\vspace{0.3cm}
\begin{center} 
\includegraphics[width=\columnwidth]{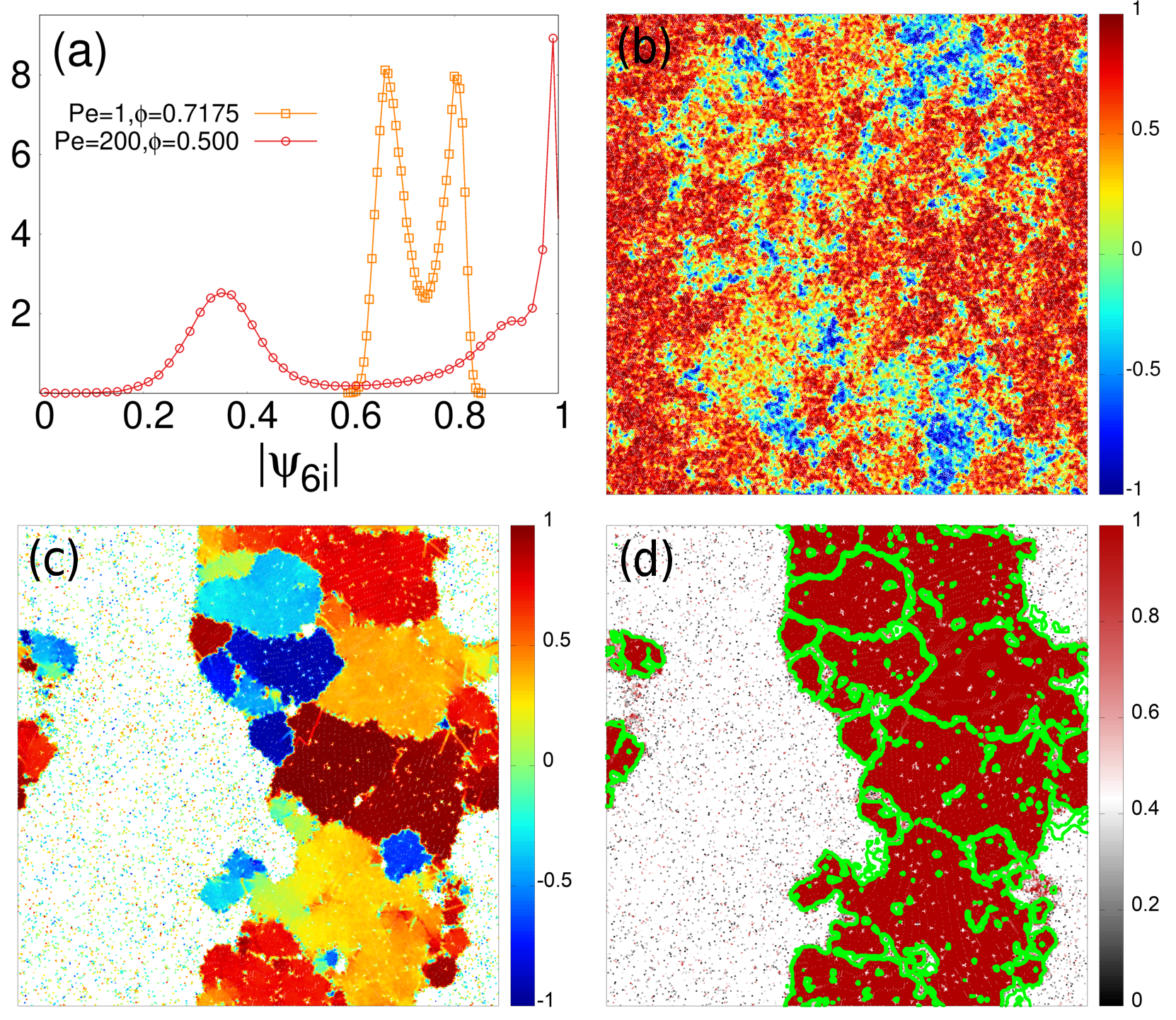}
\end{center}
\vspace{-0.25cm}
\caption{{(a) PDF of  $|\psi_{6i}|$ in the low Pe and MIPS coexistence region. (b, c) Maps of  $\psi_{6i}$, projected into the mean orientation of the system \cite{SM}, for the two cases considered in (a). (d) Snapshot showing $|\psi_{6i}|$  and the interfaces between domains with different orientational order (in green).  }
}
\label{fig:pdf-hex}
\end{figure}

{\it Orientational order and coexistence.}
%
{The maps of the local hexatic
order parameter and the PDFs of its modulus, shown in 
Fig.~\ref{fig:pdf-hex}, provide clues to understand the difference between the two sectors with phase separation at
low and high Pe. }
Close to Pe = 0 (a) the PDF is bimodal, with two peaks of roughly the same
height for this choice of parameters. The map in the insert proves the existence of a ramified but large (of the
order of the system size) region with the same local hexatic order. Under the dynamics this region changes form
but the portion of surface that it occupies remains stable. These
results are in perfect correspondence with the data for the local
densities (see the SM in~\cite{SM}). In the
MIPS region, instead, the map shows many different colors associated to
diverse local orientational ordering that do not extend over a long
distance, {even at long times}. Under the dynamics the color pattern changes considerably, with breaking and recombination of 
blocks. Differences in the maps are translated into differences in the  PDFs.
The secondary peak close to {$|\psi_{6,j} | = 0.9$} in Fig.~\ref{fig:pdf-hex} (b) is due to the interfaces between areas
with almost perfect orientational order.  
Additional maps in other sectors of the phase diagram,  PDFs of $|\psi_{6,j}|$, correlation functions and global hexatic order parameter
$\Psi=N^{-1} |\sum_j \psi_{6,j}|$ measurements are  given in the SM. 

\begin{figure}[h!]
\begin{center} 
\includegraphics[scale=0.12,angle=0]{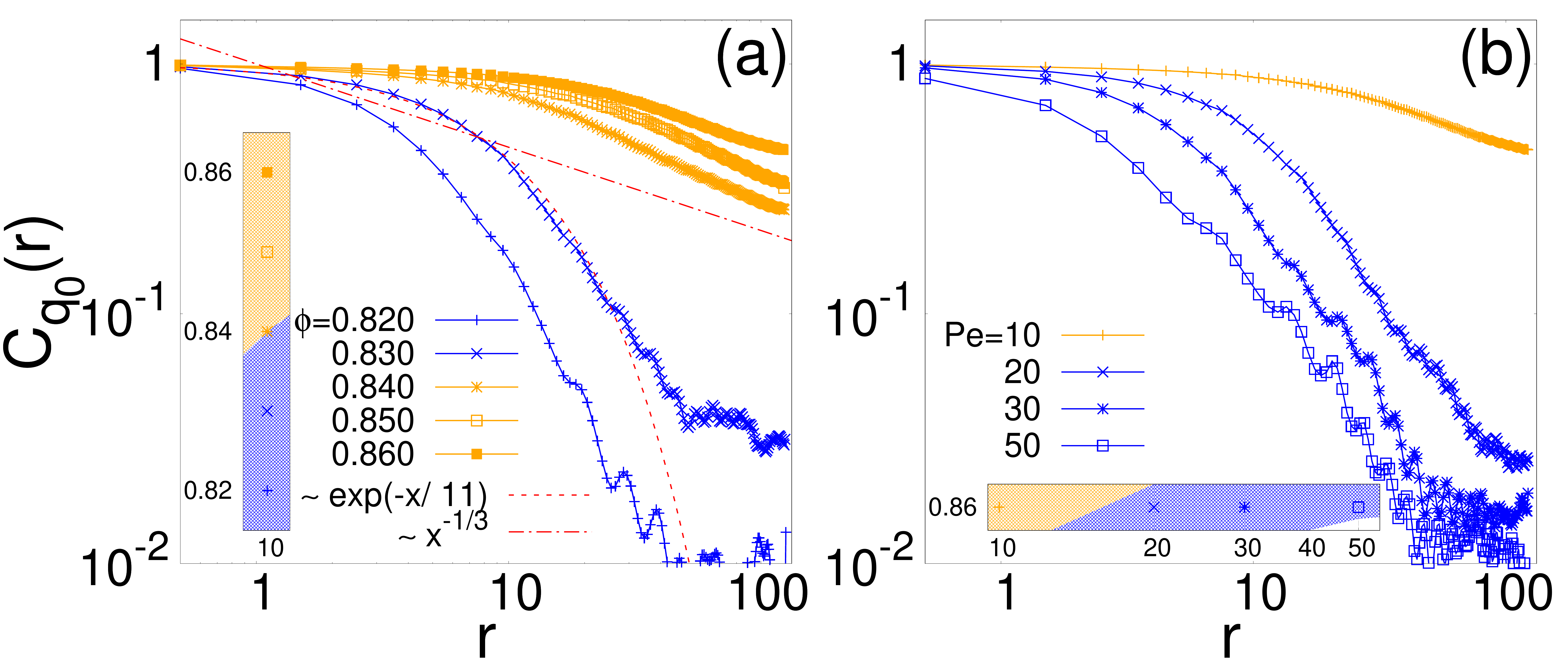}
\end{center}
\vspace{-0.35cm}
\caption{Positional correlations $C_{\bold{q}_0} (r) $ close to the hexatic-solid 
transition for several values of $\phi$ 
at fixed Pe = 10 (a) and of Pe 
at fixed $\phi=0.86$ (b). 
In (a) we  show an exponential  and an algebraic decay with power $\eta=1/3$  (corresponding to the exponent predicted by the KTHNY theory~\cite{Kosterlitz1973, Halperin1978, Young1979}). 
}
\label{fig:positional}
\end{figure}

{\it Positional order and the solid phase.}
Since it is hard to assert whether  
$g_6$ acquires long-range order or does not decay at the length-scales 
of our finite-size box, we looked for solid
quasi-long range positional order, that should be evidenced by an 
algebraic decay of 
\begin{equation}
C_{\bold{q}_0} (r) = \langle e^{{\rm i} \bold{q}_0 \cdot ({\bold r}_i - {\bold r}_j)}\rangle
\; , 
\end{equation}
 at the wave vector $\bold{q}_0$ at the maximum of the first diffraction peak of the structure factor
$S(q) = N^{-1} \sum_{i,j} e^{{\rm i} \bold{q} \cdot ({\bold r}_i - {\bold r}_j)}$. 
The change in the  $C_{\bold{q}_0}$ decay, from exponential (hexatic) to algebraic (solid) for several Pe and $\phi$, see {\it e.g.} Fig.~\ref{fig:positional},
yields the orange points in the phase diagram above which lies the solid.
Activity introduces non-equilibrium fluctuations that  destabilize order  and 
melt the solid.

Summarizing, we established the full phase diagram of Active Brownian Hard-Disks,  with 
active liquid, hexatic and solid phases, as well as coexistence and MIPS.  

First, we proved that the overall scenario of 
2D melting of passive disks is maintained for small-enough Pe. Weak  activity acts as a {perturbation} 
that destabilizes passive order, similarly to what was found in~\cite{KKK} for a system of softer disks
(no coexistence in the passive limit) evolved with Monte Carlo dynamics.
This is shown by the fact that increasing Pe both the liquid-hexatic and hexatic-solid transitions shift to higher densities,  and  
the liquid-hexatic coexistence region shrinks and eventually disappears. 
Such behavior can be due to the effective softness introduced by activity (quantified by the ratio between the 
active  and potential forces $\Gamma=\varepsilon/(\sigma_d F_{\rm act})$), since, 
{in equilibrium, particle softness reduces the  liquid-hexatic coexistence region and eventually distroys it, rendering the hexatic-liquid transition continuous} \cite{Kapfer2015b}.

At high Pe, the MIPS  region opens up {on top of} the hexatic and solid transition lines  {(differently from what was shown in~\cite{KKK})}
and   prevails the emergence of hexatic and solid order. In most of the MIPS region, 
many finite-size patches with different hexatic order coexist at any moment, but the large activity makes {them} 
regularly rearrange {\it via} breaking and recombination, very differently from what happens at low Pe. 
Above the point at which  the hexatic transition line crosses the MIPS binodal, activity triggers 
phase separation between a  low-density gas and a high-density hexatic, or solid,  {at higher Pe}. 

The discontinuity between the coexistence regions for Active Brownian Disks is distinct from what was found for active dumbbells,  for which the
large Pe phase separation was continuously connected to the zero Pe one. This difference could be due to the fact that dumbbells have a
non-convex geometry that eases jamming and the formation of local orientational order. It would be interesting to study {systems made of } 
elements that interpolate between the disk and dumbbell  geometries, and see how the topology of the phase diagram transforms from 
the one in Fig.~\ref{fig:phase-diagram} to the one in~\cite{Cugliandolo2017}.

To conclude, 
our results provide a firm basis to rationalize the phase behavior of dense active matter and understand how self-propulsion affects the 
liquid and solid phases of matter on general grounds. {The scenario we established here could be experimentally tested in, for instance, monolayers of self-propelled Janus colloids.}

\noindent
{\it Acknowledgments.}
This work was possible thanks to the access to the MareNostrum
Supercomputer at the Barcelona Supercomputing Center (BSC), 
IBM Nextscale GALILEO at CINECA
(Project  INF16-fieldturb)  under  CINECA-INFN  agreement
and  Bari ReCaS e-Infrastructure funded by MIUR through
PON  Research  and  Competitiveness  2007-2013  Call  254
Action   I. DL and IP acknowledge funding from the EU's Horizon 2020 programme under the Marie Sklodowska-Curie (IF) grant agreement No 657517
and MINECO and DURSI under projects FIS2015-67837-P and 2017SGR-884, respectively. 
LFC is a member of Institut Universitaire de France,  thanks the KITP University of 
Santa Barbara for hospitality and L. Berthier, 
P. Choudhuri, C. Dasgupta, M. Dijkstra and J. Klamser for useful discussions.


\pagebreak
\widetext
\begin{center}
 \textbf{\large Supplementary Material for \\
 ``Full phase diagram of active Brownian disks: \\
 from melting to motility-induced phase separation''}
 \date{\today}
\end{center}

\setcounter{equation}{0}
\setcounter{figure}{0}
\setcounter{table}{0}
\setcounter{page}{1}
\makeatletter
\renewcommand{\theequation}{S\arabic{equation}}
\renewcommand{\thesection}{S\arabic{section}}
\renewcommand{\thefigure}{S\arabic{figure}}

In this Supplemental Material (SM) we display further evidence for the various phases and transitions 
explained in the main text. We organize the SM in two sections in which we expand the analysis  of 
four observables: the local density in Sec.~\ref{sm:density}, and the local hexatic order parameter, hexatic correlation functions and global 
hexatic order parameter in Sec.~\ref{sm:hexatic}. We discuss their behavior and implications.


\section{The local surface fraction}
\label{sm:density}
 
We sampled the local surface fraction $\phi_i$  in the following way. 
We first divided the system in square boxes (or cells) of linear size $\sigma$. We then calculated a  coarse-grained local density 
$\phi_i$ associated to each cell $i$ by computing the mean density 
over a circle of radius $R$ centered at each cell. From the statistics of these local $\phi_i$ values we constructed a probability 
distribution function (PDF). The  choice  of the specific value of  $R$   depends on the point in the Pe-$\phi$ plane under 
consideration; indeed, we adapted  the coarse-graining to the heterogeneities of the system. Except for the simulations at Pe $=1$, 
for which we used $R=20$, the choice $R=5$ was used in all other cases.
	
\begin{figure}[h!]
	\vspace{0.5cm}
	\begin{center}
		\includegraphics[width=0.8\columnwidth]{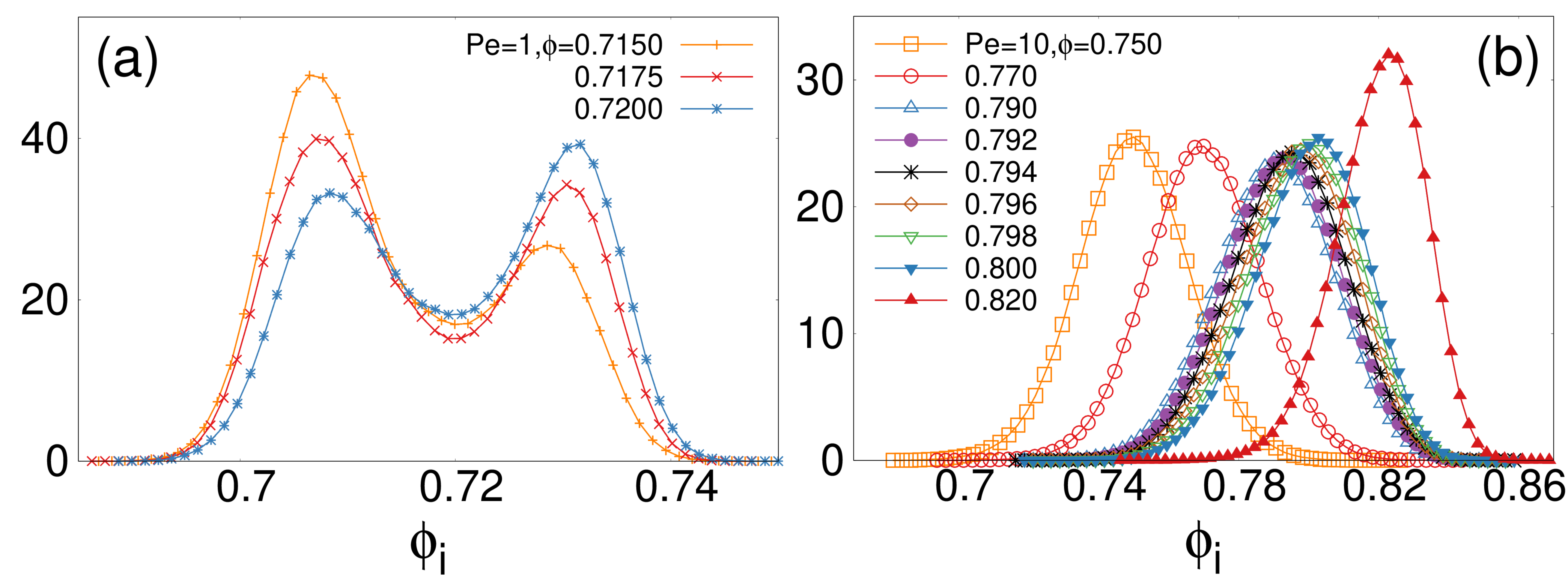}
	\end{center}
	\vspace{-0.3cm}
	\caption{Density PDFs across the liquid-hexatic coexistence region for Pe $=1$ (a) and 
	the liquid-hexatic transition  for Pe $=10$ (b). 
	Values of the global packing fraction for each curve are provided in the two keys. The coarse-graining 
	parameter $R$ was chosen to be $R=20$ in (a) and $R=5$ in (b). For Pe~$=1$ there is liquid-hexatic coexistence in between 
	$\phi\approx0.7125$ and $\phi\approx0.7275$, followed by the hexatic-solid transition at $\phi_{\rm hex-sol}\approx0.730$.
	For  Pe = 10 (no coexistence) the critical densities are $\phi_{\rm liq-hex}\approx0.795$ and $\phi_{\rm hex-sol}\approx0.840$.}
	\label{fig:PDF1}
\end{figure}

\begin{figure}[h!]
	\vspace{0.3cm}
	\begin{center}	
		\includegraphics[width=0.8\columnwidth]{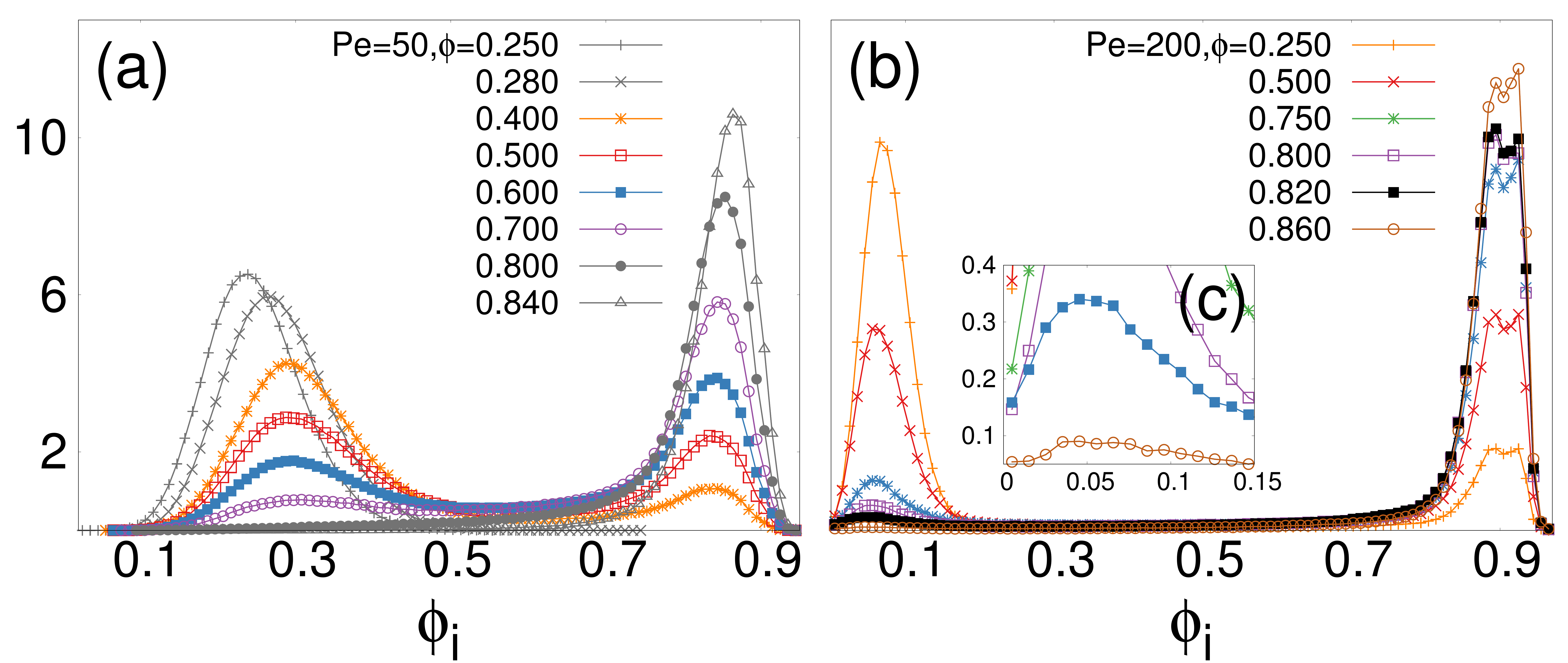}
	\end{center}
	\vspace{-0.3cm}
	\caption{Density PDFs across the MIPS region for Pe $=50$ (a) and Pe $=200$ (b). 
	Curves with a single mode are shown in grey (uniform phase), while bimodal curves are shown with 
	colors (phase coexistence). Values of the global packing fraction for each curve are provided in the two keys and the coarse-graining parameter 
	$R=5$ was used in both panels.
	As a references, for Pe = 50 we observed MIPS for $0.310\lesssim\phi\lesssim0.800$, and we measured $\phi_{\rm liq-hex}\approx0.855$ and 
	$\phi_{\rm hex-sol}\approx0.890$
	from the change from exponential to algebraic decay of the  hexatic and positional correlation functions, respectively. 
	For Pe = 200, MIPS was found for  $0.100\lesssim\phi\lesssim0.900$ and we measured $\phi_{\rm liq-hex}\approx0.860$ and $\phi_{\rm hex-sol}
	\approx0.890$.
	}
	\label{fig:PDF3}
\end{figure}

Representative local density PDFs are shown in Figs.~\ref{fig:PDF1} and \ref{fig:PDF3}, where data for Pe = 1, Pe = 10, Pe = 50 and 
Pe = 200 are plotted. The first case corresponds to  the low activity limit and the global packing fractions are chosen so that 
they lie within the black coexistence region in the phase diagram in Fig. 1 of the main text. The second case corresponds to the intermediate Pe region where no coexistence is observed, and the packing fractions are chosen, in particular, in the vicinity of the liquid-hexatic transition. The latter two cases are, instead, beyond the 
critical point towards MIPS. 

The local density PDFs provide further evidence for phase coexistence both in the low-Pe 
(hexatic-liquid coexistence) and high-Pe (MIPS) regimes (see the phase diagram in Fig.~1 in the main text). 
In the low Pe limit, Fig.~\ref{fig:PDF1} (a), 
we see that the three curves displayed, corresponding to global densities varying over the very narrow interval  [0.715 : 0.72], are bimodal with the 
weight under the two peaks slowly transferring from the one at low $\phi_i$ to the one at high $\phi_i$ for increasing $\phi$. 
For higher Pe values  we plot curves for a larger range of variation of $\phi$. No two peak structure
is found beyond Pe $\simeq 3$, that is to say, beyond the ending point on the black region of the phase diagram at low 
Pe values, and before the critical point for MIPS, the second black region in the phase diagram, is reached. 
Figure~\ref{fig:PDF1}~(b), where data for Pe  = 10 are plotted, demonstrates this point: for all $\phi$ the curves are bell-shaped and their average and 
typical value displace in unison towards higher values for increasing $\phi$. This proves that there is no 
coexistence for this Pe and, in fact, for a rather wide range of Pe values as shown in the phase diagram in 
Fig.~1 of the main text. Finally, we investigate what happens at large Pe values. For Pe = 50, in Fig.~\ref{fig:PDF3}~(a)  
we plot in grey the curves with just one 
peak, either at low or high values of $\phi_i$, and in color the bimodal curves that show a similar transport of weight 
from low to high local densities upon increasing $\phi$. Similar features characterize the data at Pe  = 200 depicted in panel 
(b) in the same figure. The insert, labeled (c), zooms over the low local density values showing that systems at average packing fraction $\phi=0.820$ and $\phi=0.860$ are in the coexistence region.

The open circular symbols in the phase diagram in Fig.~1 were obtained from the location of the global packing fractions that limit the 
region with a double peaked PDF of local densities, in a very satisfactory agreement with the data stemming from the pressure measurements 
shown with filled black dots. 

We note that from local density measurements we cannot investigate whether the dense phase has orientational or positional  
order. We addressed this point with local hexatic order parameter computations and position correlation functions, respectively.
 
 \section{The hexatic order}
 \label{sm:hexatic}

As discussed in the main text, we calculated the local hexatic order parameter 
\begin{equation}
	\label{eq:local_hexatic}
	\psi_{6,j}=\psi_6({\mathbf r}_j) = N_{ j}^{-1} \sum_{k=1}^{N_{ j}} e^{{\rm i} \theta_{jk}}
\end{equation}
for each disk in the system, where $N_j$ are the first neighbours of the disk $j$. In order to do so, we  built up the 
nearest-neighbors network by means of a 
Voronoi tessellation.  We studied the local orientational order in the system constructing maps of this order parameter
and its PDF. With this quantity we also computed correlation 
functions and we defined a global order parameter. We discuss all these measurements in this Section.

\subsection{Maps of local hexatic order}

\begin{figure}[h!]
	\vspace{0.3cm}
	\begin{center}
	\includegraphics[width=\columnwidth]{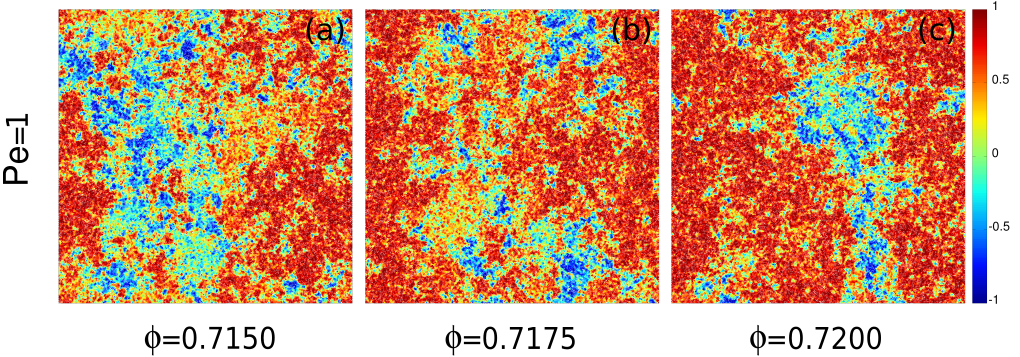}
	\end{center}
	\caption{\textbf{Active liquid-hexatic coexistence.} Maps of the local hexatic order parameter, $\psi_{6,j}$, in 
	the stationary regime  for Pe $=1$ 
	across the coexistence region at $\phi=0.715$ (a), $\phi=0.7175$ (b) and $\phi=0.72$  (c). 
 	We recall that for Pe~$=1$ we have liquid-hexatic coexistence in between 
	$\phi\approx0.7125$ and $\phi\approx0.7275$, followed by the hexatic-solid transition at $\phi_{\rm hex-sol}\approx0.730$.
	}
	\label{fig:pe1}
\end{figure}

\begin{figure}[h]
	\vspace{0.3cm}
	\begin{center}
	\includegraphics[width=\columnwidth]{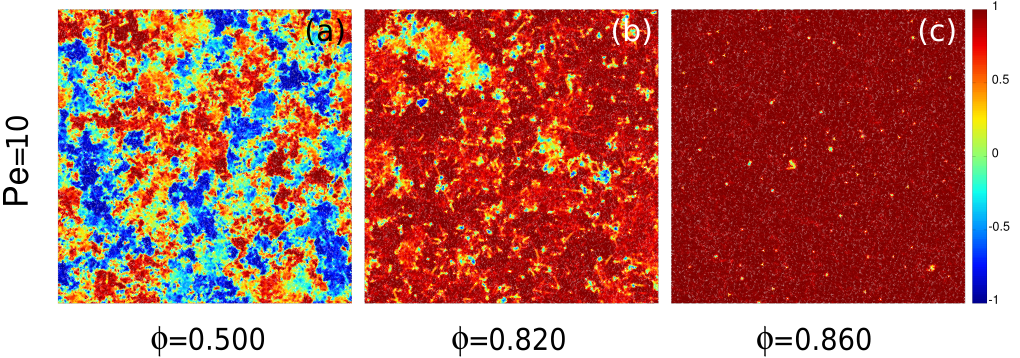}
	\end{center}
	\caption{\textbf{Active liquid, hexatic and solid phases.} Maps of the local hexatic order parameter, $\psi_{6,j}$, in the stationary 
	regime   for Pe $=10$ at $\phi=0.500$ 
	 (in the active liquid phase) (a), $\phi=0.820$ (in the active hexatic phase) (b) and $\phi=0.860$ 
	 (in the active solid phase) (c). 
	Note that at this Pe the critical densities $\phi_{\rm liq-hex}\approx0.795$ and $\phi_{\rm hex-sol}\approx0.840$ were obtained from the analysis of the 
	orientational and positional correlation functions.
	}
	\label{fig:pe10}
\end{figure}

\begin{figure}[h!]
	\vspace{0.3cm}
	\begin{center}
	\includegraphics[width=\columnwidth]{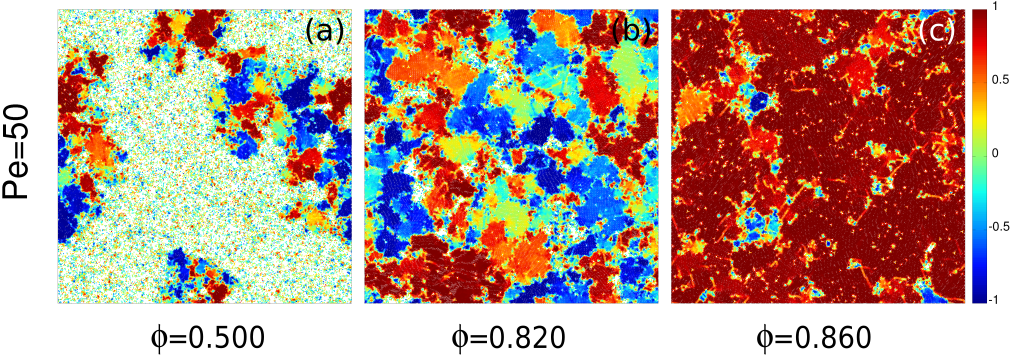}
	\end{center}
	\caption{\textbf{MIPS and active hexatic phase.} Maps of the local hexatic order parameter, $\psi_{6,j}$, 
	in the stationary regime for Pe $=50$ at $\phi=0.500$  (in the MIPS coexistence region) (a), 
	$\phi=0.820$ (in the dense active liquid phase) (b) and $\phi=0.860$ (in the active hexatic phase) (c). 
	For this value of Pe we observed MIPS for $0.310\lesssim\phi\lesssim0.800$, and we measured $\phi_{\rm liq-hex}\approx0.855$ and $\phi_{\rm hex-sol}\approx0.890$.
	}
	\label{fig:pe50}
\end{figure}

\begin{figure}[h]
	\vspace{0.3cm}
	\begin{center}
	\includegraphics[width=\columnwidth]{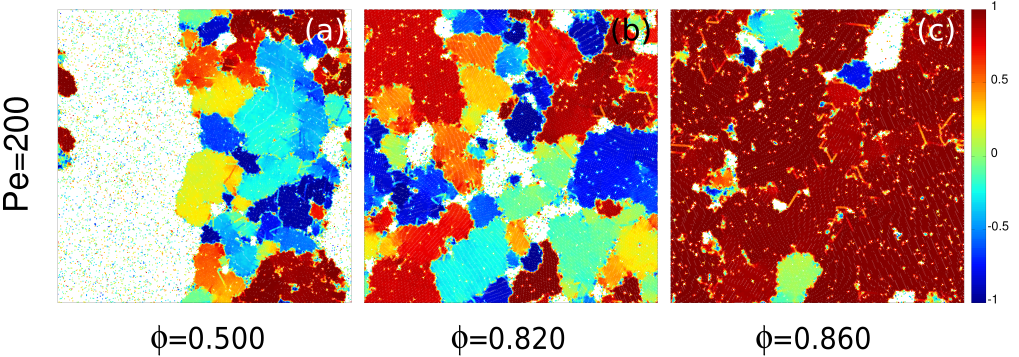}
	\end{center}
	\caption{
	\textbf{Dilute-dense coexistence by MIPS.}  Stationary-state snapshots showing the local hexatic order parameter 
	$\psi_{6,j}$ for Pe $=200$ within the MIPS coexistence region at different packing fractions: $\phi=0.500$ (a),  
	$\phi=0.820$  (b) and  $\phi=0.860$ (c). At Pe = 200, and for high enough density,  activity induces phase 
	separation into a low density gas and a high density solid phase.
	Note, as a reference, that for this Pe we observed MIPS for 
	$0.100\lesssim\phi\lesssim0.900$, and we measured $\phi_{\rm liq-hex}\approx0.860$ and $\phi_{\rm hex-sol}
	\approx0.890$ from the change from exponential to algebraic decay of the  hexatic and positional correlation functions, respectively. 	
	}
	\label{fig:pe200}
\end{figure}

In the following figures we visualize the local hexatic order using the method proposed in~\cite{Bernard2011g}: first, we 
project the complex local values $\psi_{6,j}$ onto the direction of the mean orientation $N^{-1}  \sum_{i=1}^N \psi_{6,i}$, where the 
sum runs over all particles in the sample. 
We then associate a color code to each bead according to this normalized projection. Regions with orientational order have 
uniform color. The dominating orientational ordering is  painted in dark red, and the hierarchy follows the scale shown at the 
extreme right of the panels in Fig.~\ref{fig:pe1}, \ref{fig:pe10}, \ref{fig:pe50} and \ref{fig:pe200}. 
Due to the six-fold symmetry of the ordered state, 
blue regions are hexatically ordered along a lattice which is rotated by $\pi/2$ with respect to the one of the dark red regions. 
Green spots, which correspond to zero local hexatic order parameter in the color code, represent 
regions rotated by $\pi/4$ from the red ones. 

The information stemming from the PDFs of local densities in Figs.~\ref{fig:PDF1} and \ref{fig:PDF3} is complemented by 
Figs.~\ref{fig:pe1},~\ref{fig:pe10}, \ref{fig:pe50} and \ref{fig:pe200}, where we show some representative snapshots of the system 
illustrating the nature of the different regimes reported in the main text. The snapshots encode the maps of local 
hexatic order, as explained in the previous paragraph. All the pictures correspond to the state of the 
system after letting it relax for about $10^6$ MDs  from a fully random initial condition.

In the low Pe case displayed in Fig.~\ref{fig:pe1}, where the three panels span the coexistence region of the phase diagram, 
we see large (non compact) regions with red color that correspond to the same local hexatic order that are surrounded in a 
rather disordered way by regions with no global hexatic order. These features are very similar to the ones seen in the active 
dumbbell system studied in~\cite{cugliandolo2017} and in the passive disk models with sufficiently hard repulsive potential  
studied by Krauth and collaborators~\cite{Bernard2011g,Kapfer2015b,Engel2015}.

In Fig.~\ref{fig:pe10} we display maps at Pe $=10$, an activity for which we do not see coexistence,  and the transition pattern is simpler with one line separating active liquid and 
active hexatic phases and another one separating the active hexatic from the active solid phase.  To start with, the color map in panel (a) looks different from the 
ones shown in the three panels in Fig.~\ref{fig:pe1}, with no large red zone in this configuration, that  corresponds to a low global 
packing fraction  ($\phi=0.5$) and that  we interpret as an active liquid one. The intermediate panel (b), obtained for $\phi=0.820$, displays  an almost uniform, relatively light, red pattern 
 and it lies in the active hexatic phase. Finally, a much darker uniform red state is shown in the last panel (c), in which $\phi=0.860$, and the system is an active 
solid, as confirmed by the analysis of the positional correlation functions shown in the main text.

The case Pe $=50$ is above the critical Pe value for which MIPS occurs. In the phase diagram in Fig.~1 in the main text, 
one can see a reentrance of the active liquid above the MIPS region before entering the hexatic phase at an even higher global packing fraction.
Typical color maps of the local hexatic order parameter are shown in Fig.~\ref{fig:pe50} for global densities that lie well within the MIPS region
(a), in the reentrant active liquid (b) and in the active hexatic (c). It is hard to assess from the snapshot in panel (b) what is the 
nature of the system  for these parameters, although it is clear that there is no predominant hexatic order in the sample. 
The conclusion about reentrance was drawn from the analysis of other observables, 
notably, the pressure and correlation functions.

At Pe $=200$, the case considered in Fig. \ref{fig:pe200}, MIPS generates the de-mixing of the system into a very low-density  
($\phi_{\rm low}\approx 0.07$) and a rather high-density ($\phi_{\rm high}\approx 0.9$) phase with, on top, local hexatic order of different kinds. 
This is illustrated in the snapshots in panels (a) and (b) by the dense regions of particles sharing the same color, in coexistence with very 
low density regions. As largely reported in the MIPS literature \cite{Fily2012}, the dense phase induced by activity  is subjected to anomalously large density fluctuations: it continuously breaks and reforms, giving rise to the observed distribution of orientationally ordered 
patches of finite size (instead of a single uniform one). We confirmed this claim by following the time evolution of these states (not shown). 
Panel (a) is also shown as an inset in Fig.~5 in the main text. Next to it, another inset explains the pattern of domains with 
different orientational order: the map of $|\psi_{6,i}|$ is shown with a red scale and interfaces between domains with different 
$\psi_{6,i}$ are highlighted in green.
For high enough global packing fraction, $\phi=0.860$, the phase separation is 
between an active gas and an active solid, see panel (c) with a clear hole in the upper right corner.

\subsection{PDFs of the modulus of the local hexatic order parameter}

We now make the analysis of the local hexatic order quantitative by tracing the PDFs of the modulus of the 
corresponding local order parameter.

\begin{figure}[h!]
	\vspace{0.5cm}
	\begin{center}
		\includegraphics[width=0.8\columnwidth]{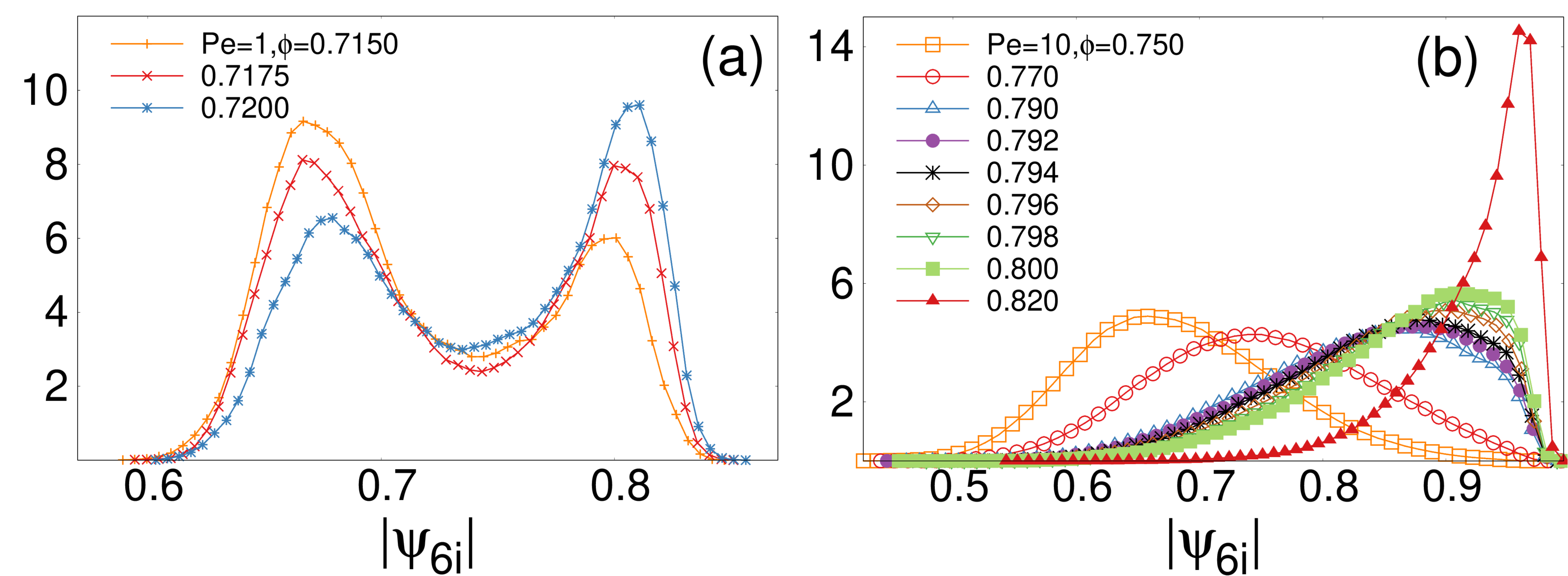}
	\end{center}
	\vspace{-0.3cm}
	\caption{PDF of the modulus of the local hexatic order parameter across the liquid-hexatic coexistence region for Pe $=1$ (a) and 
	the liquid-hexatic transition for Pe $=10$ (b). 
	Values of the global packing fraction for each curve are provided in the keys. For Pe~$=1$,  liquid-hexatic coexistence is found between 
	$\phi\approx0.7125$ and $\phi\approx0.7275$, and it is followed by the hexatic-solid transition at $\phi_{\rm hex-sol}\approx0.730$.
	For  Pe = 10 the critical densities are $\phi_{\rm liq-hex}\approx0.795$ and $\phi_{\rm hex-sol}\approx0.840$.}
	\label{fig:PDF-hex-1}
\end{figure}

At low Pe, Fig.~\ref{fig:PDF-hex-1},  the PDF has a bimodal form, in full correspondence with the one
for the local density shown in Fig.~\ref{fig:PDF1}. The peak at a high value of $|\psi_{6,j}|$  corresponds to regions with 
local hexatic order while the one at low value of $|\psi_{6,j}|$ represents the disordered regions. (We note that a disordered 
liquid phase does not have a vanishing modulus of the local hexatic order parameter but a unimodal distribution with average and typical 
 values that increase with the packing fraction.)

Increasing Pe beyond the end of the coexistence region connected to the
passive limit (end of the black region in the phase diagram connected to Pe = 0), 
the PDF of $|\psi_{6,j}|$ dramatically changes form, and has only one peak
that displaces with increasing packing fraction from a low to a high value
of $|\psi_{6,j}|$ but never develops a double peak structure.  These facts are shown in 
Fig.~\ref{fig:PDF-hex-1}. Qualitatively, the form of the PDF is the same as the one for the 
local density shown in Fig.~\ref{fig:PDF1} (b). Therefore, from this measurement we cannot locate the 
phase transition towards the phase with hexatic order. This has to be done from the analysis of the 
orientational correlation functions and the Binder parameter, as explained in the main text.

\begin{figure}[h!]
	\vspace{0.3cm}
	\begin{center}	
		\includegraphics[width=0.8\columnwidth]{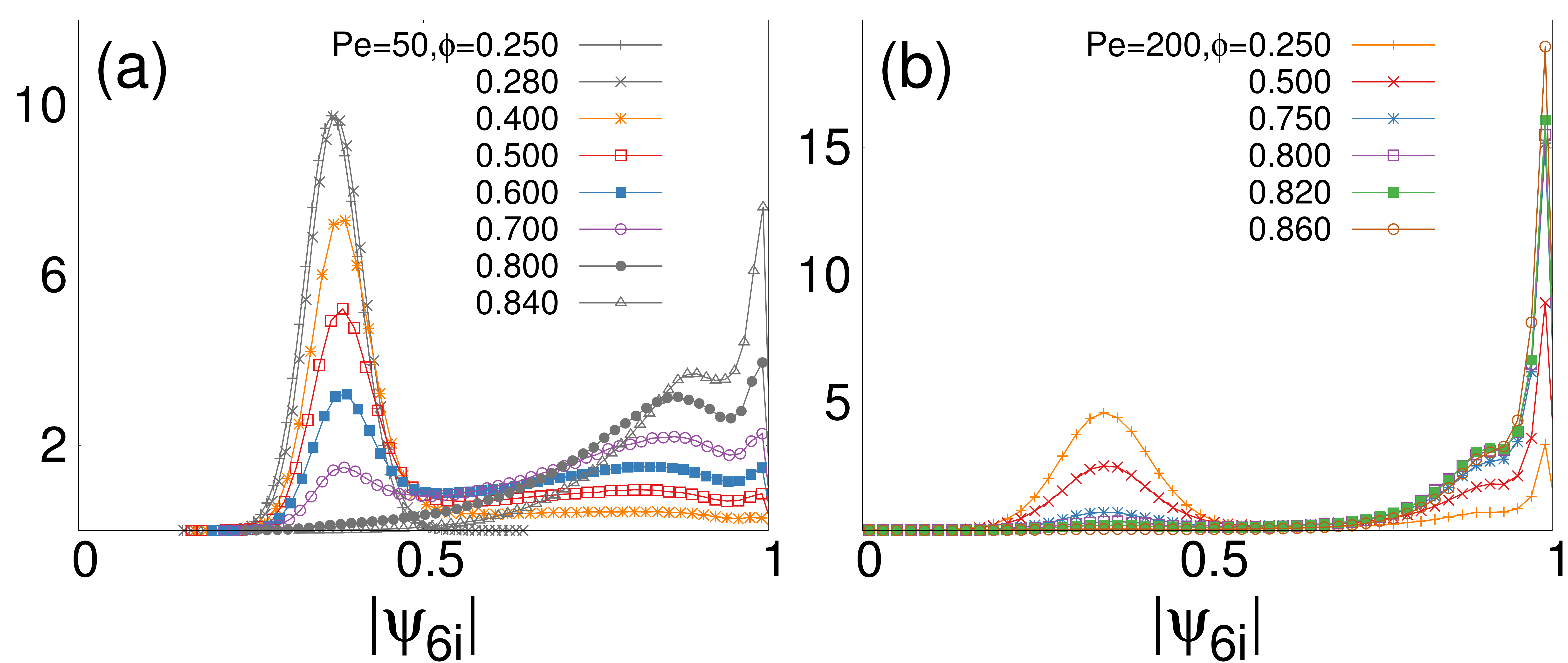}
	\end{center}
	\vspace{-0.3cm}
	\caption{PDF of the modulus of the local hexatic order parameter across the MIPS region for Pe $=50$ in (a) and Pe $=200$ in (b). 
	Curves with a single mode are shown in grey (uniform phase), while bimodal curves are shown with 
	colors (phase coexistence). Values of the global packing fraction for each curve are provided in the key. 
	For Pe = 50 we observed MIPS for $0.310\lesssim\phi\lesssim0.800$, and we measured $\phi_{\rm liq-hex}\approx0.855$ and 
	$\phi_{\rm hex-sol}\approx0.890$ while 
	for Pe = 200, MIPS was found for  $0.100\lesssim\phi\lesssim0.900$ and we measured $\phi_{\rm liq-hex}\approx0.860$ and $\phi_{\rm hex-sol}
	\approx0.890$. See the text for a discussion of the 
	form of these curves, especially the secondary peak appearing below the one close to $|\psi_{6,j}|=1$.}
	\label{fig:PDF-hex-3}
\end{figure}

Finally, inside the MIPS region, the PDF of the modulus of the local
hexatic order parameter acquires, again, a multi peak structure, see Fig.~\ref{fig:PDF-hex-3}. The one at
low values of $|\psi_{6,j}|$ corresponds to the disordered and in some cases
almost empty regions. Note that, basically,  the position of the maximum of this peak does not depend
 on $\phi$ within the scale of this figure and it coincides with the one of the homogeneous phase at very low $\phi$. 
 The peak at $|\psi_{6,j}|$ close to one
indicates that there are regions in the system with almost perfect local
hexatic order. Below this peak appears a second one, of lesser height,
that is associated to the disks that are close to the interfaces of the 
perfectly ordered domains (of finite size). This fact is proven by the map in the second insert in  
panel (b) of Fig.~5 in the main text. This figure is to be compared to Fig.~\ref{fig:PDF3}
where the PDFs of local density for the same parameters are shown.

\subsection{Hexatic correlation function}

Data for the correlation function of the local hexatic order parameter measured a distant points in space, 
\begin{equation}
g_6(r) = \left. \langle \psi^*_{6,j} \psi_{6,k} \rangle\right|_{|{\mathbf r}_j - {\mathbf r}_k| = r} / \langle |\psi_{6,j}|^2 \rangle
\; , 
\end{equation}
at Pe = 10 and various global packing fractions $\phi$, and various Pe and $\phi=0.8$, were shown in Fig. 3 in the main text.  The change from 
exponential to algebraic was used as a criterium to locate the transition from the active  liquid to  the active hexatic phases.
Here we exhibit the behavior of these correlation functions at a higher value of Pe where the change operates within the 
MIPS region and informs us about the nature of the dense phase that has either short, quasi or proper long range
hexatic order. 

\begin{figure}[h!]
	\vspace{0.3cm}
	\begin{center}	
		\includegraphics[width=0.8\columnwidth]{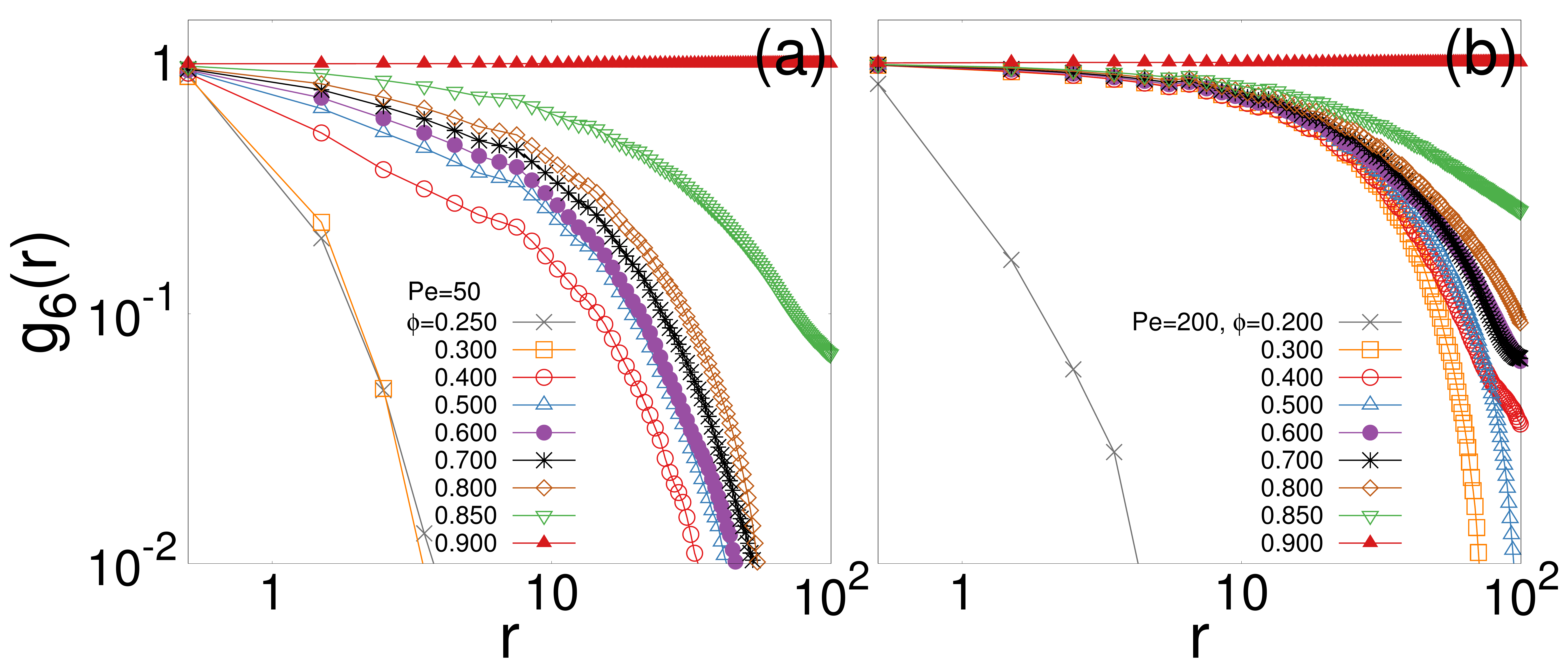}
	\end{center}
	\vspace{-0.3cm}
	\caption{Distance dependence of the hexatic order correlation function $g_6(r)$ at high Pe, Pe = 50 (a) and Pe = 200 (b) for different 
	global packing fractions given in the keys. We recall 
	$\phi_{\rm liq-hex} \approx 0.855$ and $\phi_{\rm hex-sol} \approx 0.890$ for Pe = 50 and 
	$\phi_{\rm liq-hex} \approx 0.860$ and $\phi_{\rm hex-sol} \approx 0.890$ for Pe = 200. 
	The black curves are therefore very close to the limit towards MIPS with a dense-hexatic phase and the green one is within the solid phase.}
	\label{fig:corr-hex}
\end{figure}

In Fig.~\ref{fig:corr-hex} we plot the correlations of the hexatic order parameter at Pe = 50 in (a) and Pe = 200 in (b), and we follow its
form  for varying $\phi$. Differently from what was shown for the coexistence region at low Pe, for most densities the correlations decay very fast, as 
exponentials,  where MIPS is between a dilute and a dense phase but, clearly, the latter has no long-range nor quasi long-range 
orientational order. Two curves in both panels do not decay so fast:  the black and green sets of data. Concerning the 
black curve in (a), it was obtained for a global density that falls, according to our measurements, in the reentrant active liquid phase and it is still 
exponential. The black curve in (b) is within the MIPS area but close to the line beyond which the separation involves a dense phase with hexatic order.
The green curves instead correspond to a global density that is very close to the one of close packing and the
system behaves as a solid in these length scales.

\subsection{Global hexatic order}
\label{sec:three-SM}

\begin{figure}[h]
\vspace{0.35cm}
	\begin{center}
	\includegraphics[width=0.8\columnwidth]{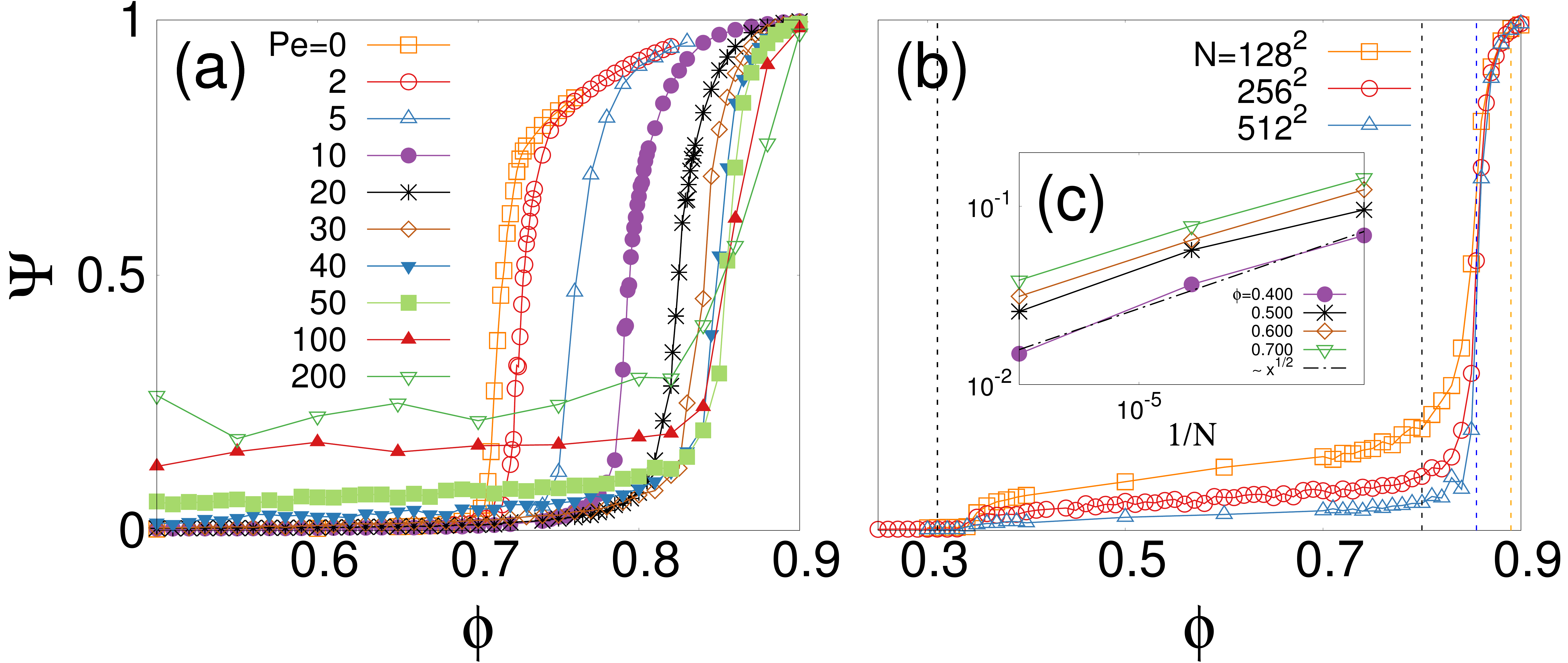}
	\end{center}
	\caption{Global hexatic parameter, defined in~Eq.~(\ref{eq:global_hexatic}), as a function of the global packing fraction. 
	(a) The curves were obtained for different values of Pe given in the key and $N=256^2$.
	(b) In all cases Pe = 50 and different number of particles given in the key. 
	In the right panel, the short straight dashed lines indicate the values of $\phi$ that limit MIPS, the liquid-hexatic phase transition (blue)
	and the hexatic-solid transition (orange) for this Pe value. In the inset (c) the finite $N$ behaviour of the value of $\Psi$ for various 
	$\phi$ is analyzed. The dashed dotted line, is simply proportional to $N^{-1/2}$, showing that a vanishing infinite size limit is 
	compatible with the data.
	}
	\label{fig:global_hexatic}
\end{figure}

We used the local hexatic parameter to compute the global hexatic order parameter
\begin{equation}
	\label{eq:global_hexatic}
	\Psi = \frac{1}{N} \left| \sum_{i=1}^N \psi_{6,i} \right| 
	\; , 
\end{equation}
that we show in Fig.~\ref{fig:global_hexatic}, as a function of $\phi$ for 
several Pe values and system sizes. In panel (a), where we used a system with $N=256^2$ particles, 
for each Pe, the global hexatic order  parameter displays a sharp  increase around a value of $\phi$
that increases with increasing activity. This shows that activity shifts the emergence of hexatic order toward higher densities. 
The position of the critical curve (Pe, $\phi$) separating the active liquid ($\Psi=0$) from the other phases with 
various criteria yield values that are in agreement with the determination of the 
hexatic transition line from the orientational correlations and the Binder cumulant (see the main text).

The curves in (a) retain roughly the same form for all Pe values until, say, Pe = 40 where  we see that they detach from zero at small 
densities developing a kind of plateau, the height of which increases with increasing activity. The way in which the highest value $\Psi=1$ is 
reached in the ramping part of the curves $\Psi(\phi)$ is also modified at high Pe. In panel (b) we have checked how do the curves for fixed Pe = 50 depend on the 
number of particles in the system. From the main panel one sees that the form of the curves gets closer to the one of the weaker activity 
systems for increasing system size. We have also indicated with vertical dotted lines with different color the  important density values:
the limits of MIPS in black, the transition to the hexatic phase (as obtained from the correlation functions of $\psi_{6,j}$) in blue and the 
one to the solid (as obtained from the positional correlation functions) in orange.
In the inset in panel (b) we study the infinite $N$ limit of the plateau value. We make  four choices of the packing fraction $\phi =
0.4, 0.5, 0.6, 0.7$, at which we trace the value of $\Psi$ against $1/N$. Although we only have three system sizes to work with, 
and it is not possible to conclude on the finite $N$ dependence and $N\to\infty$ limit of the data beyond any doubt, a na\"{\i}ve comparison to the 
law $1/N^{1/2}$ is quite acceptable. This suggests that, in the infinite size limit, the global order parameter vanishes in the 
MIPS region below the (blue dotted) line in the phase diagram that indicates the entrance into coexistence with hexatic order.

\section{Number fluctuations}
\label{sec:four-SM}
\begin{figure}[h]
\vspace{0.35cm}
	\begin{center}
	\includegraphics[width=0.4\columnwidth]{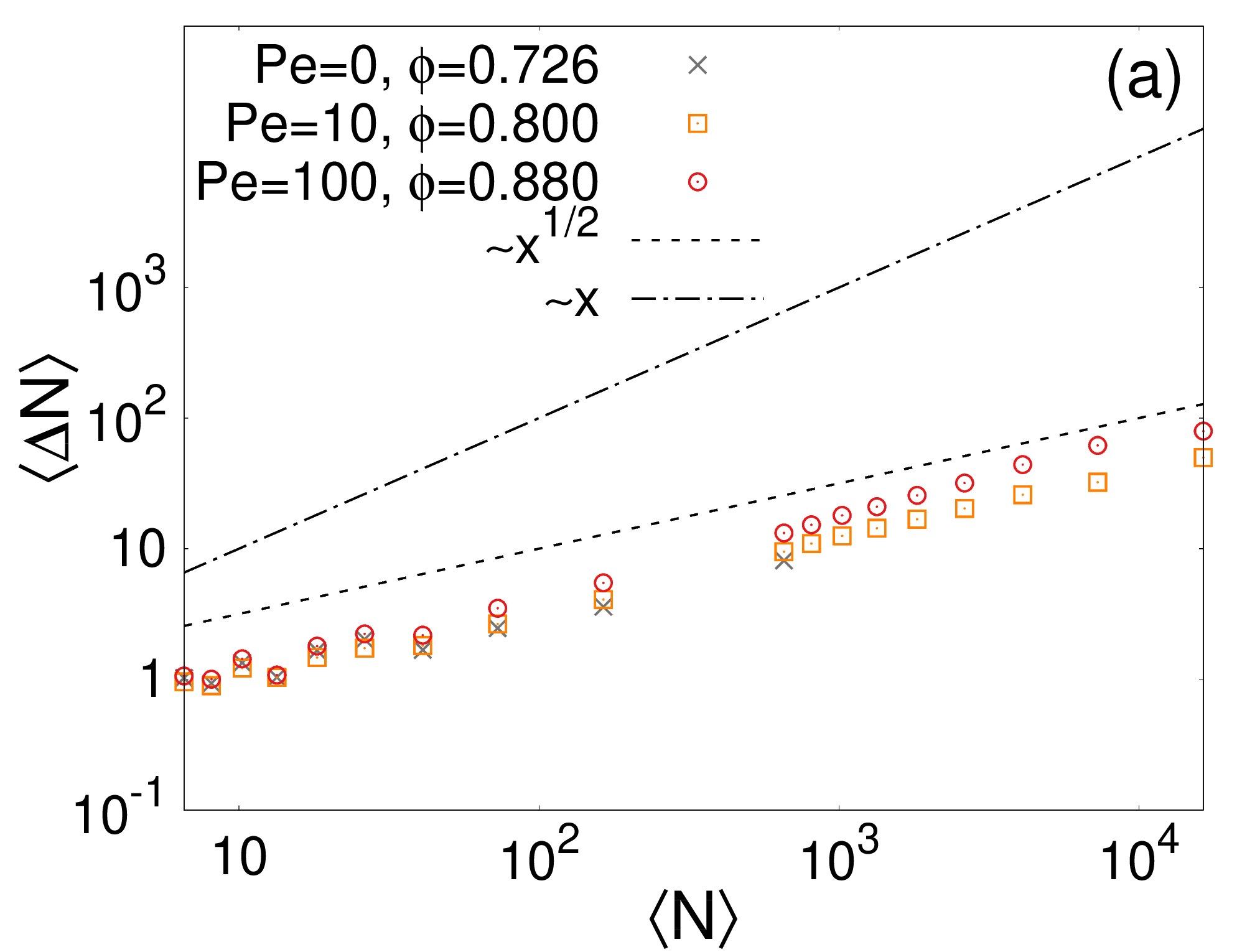}\includegraphics[width=0.4\columnwidth]{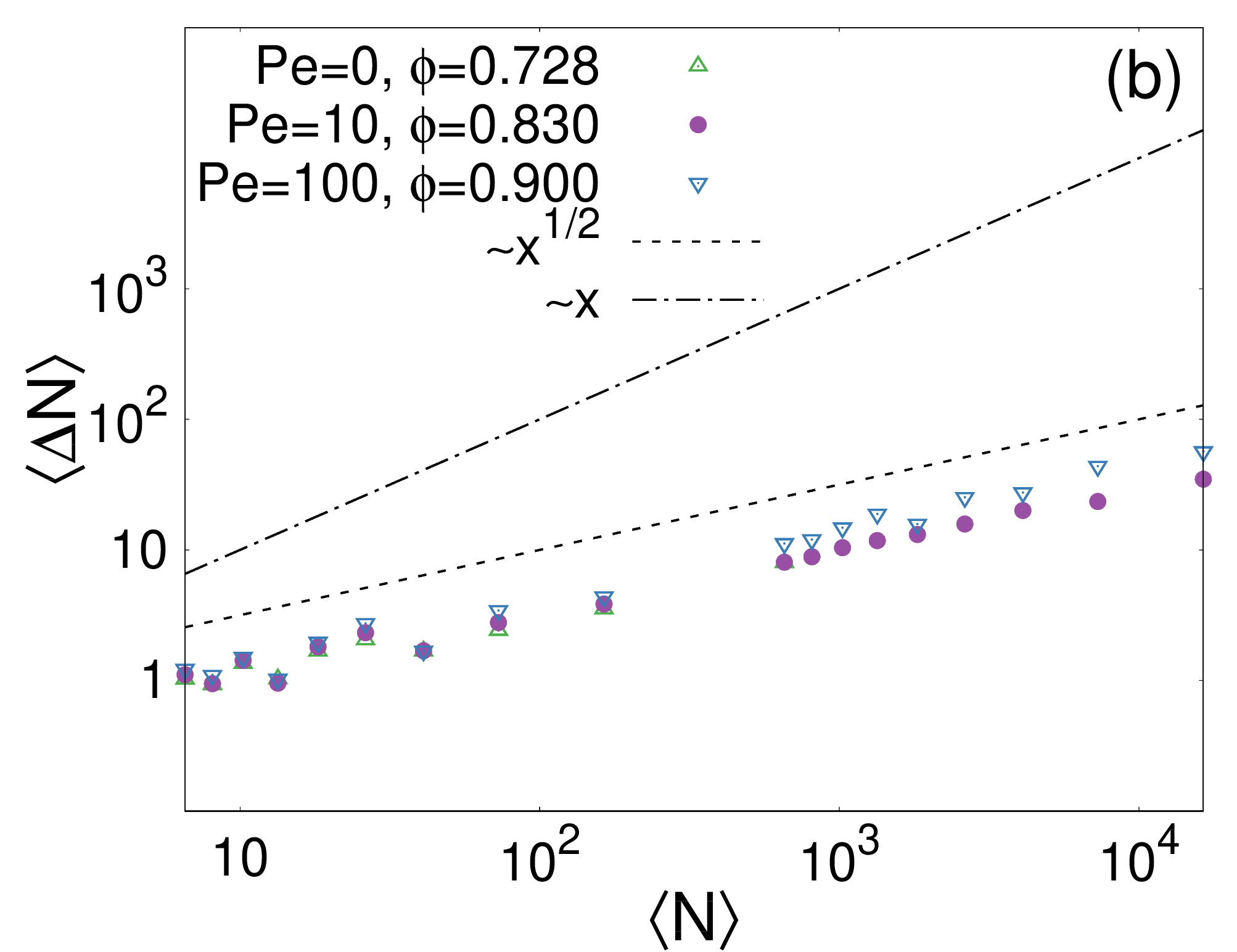}
	\includegraphics[width=0.4\columnwidth]{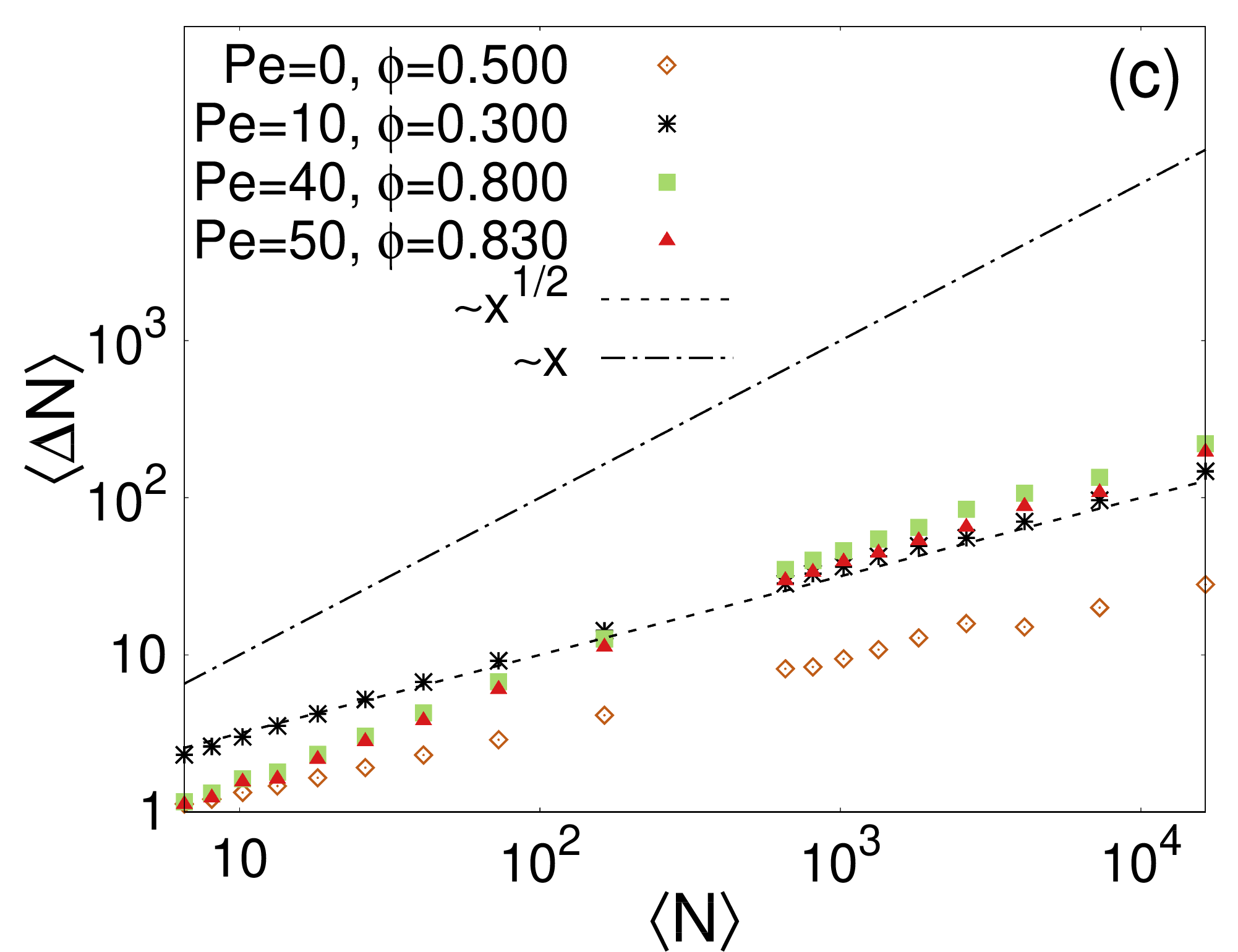}\includegraphics[width=0.4\columnwidth]{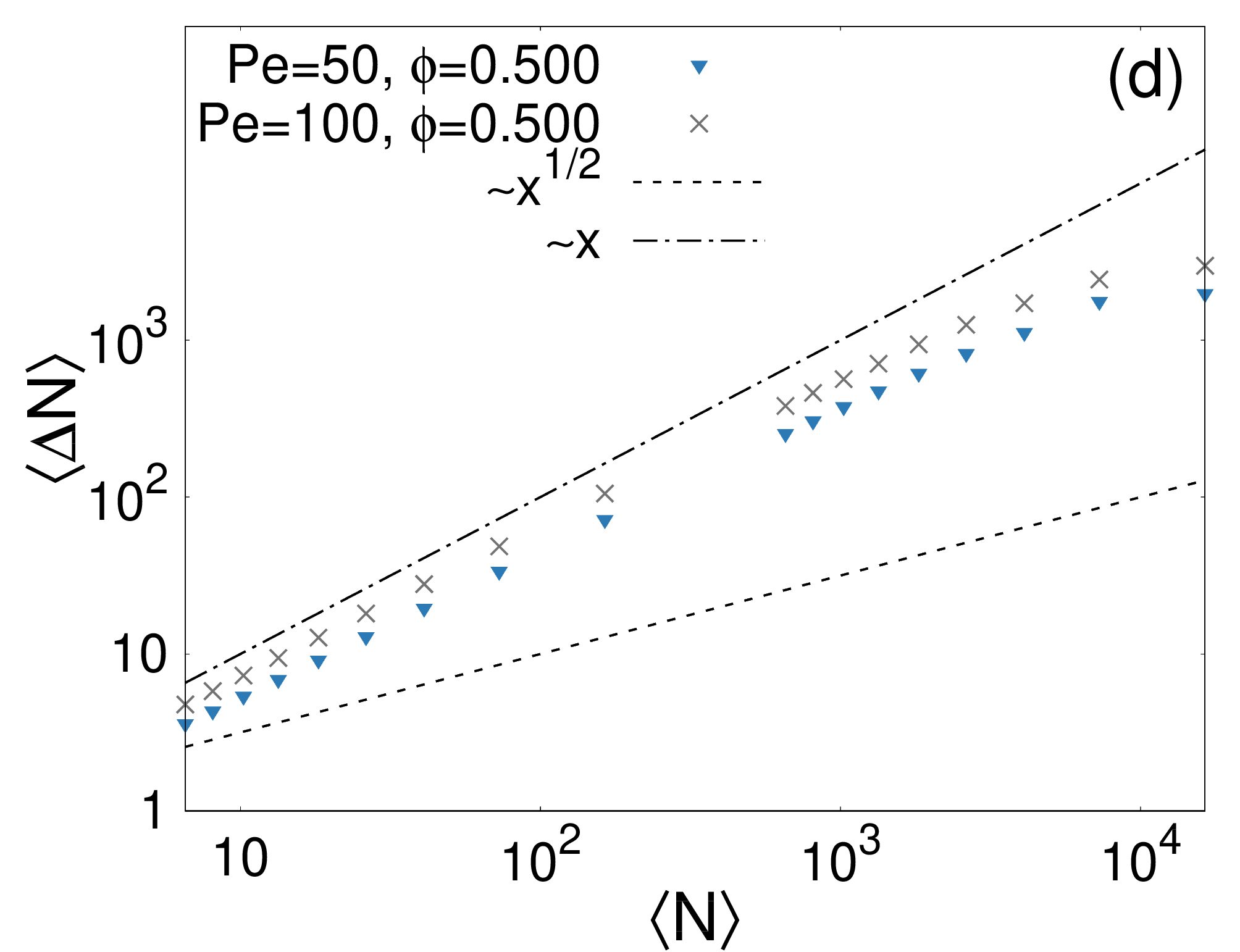}
	\end{center}
	\caption{Number fluctuations $\langle \Delta N\rangle$ vs. $\langle N\rangle$ in the hexatic (a), solid (b), liquid (c) and MIPS coexistence region (d) in log-log scale. We also shown for comparison $\langle \Delta N\rangle\propto \sqrt{\langle N\rangle}$ and $\langle \Delta N\rangle\propto \langle N\rangle$ in broken lines.
	}
	\label{fig:GNF}
\end{figure}

We provide below some simulation results of the number fluctuations in the different regimes we identified. These are calculated taking averages over system sub-boxes of different sizes and computing the deviation of the number of particles inside $\langle \Delta N\rangle$ and its mean $\langle N\rangle$.  
The results are shown in Fig. \ref{fig:GNF} for representative cases in the hexatic (a), solid (b), liquid (c) and MIPS coexistence region (d). Anomalously large fluctuations (i.e. $\langle \Delta N\rangle\sim \langle N\rangle^{\alpha}$ with $\alpha>1/2$) are observed in the MIPS coexistence regime only, as it as been previously reported \cite{Fily2012}. 

\end{document}